\documentclass[12pt]{article}                 

\newlength{\abstractwidth}
\usepackage{epsf}
\usepackage{psfig}

\renewcommand{\thefootnote}{\fnsymbol{footnote}}
\renewcommand{\thanks}[1]{\footnote{#1}} 
\newcommand{\starttext}{
\setcounter{footnote}{0}
\renewcommand{\thefootnote}{\arabic{footnote}}}

\newcommand{\bea}{\begin{eqnarray}}
\newcommand{\eea}{\end{eqnarray}}
\newcommand{\ee}{\end{equation}}
\newcommand{\be}{\begin{equation}}

\def\B{{\cal B}}
\def\C{{\cal C}}

\def\L{{\cal L}}

\def\N{{\cal N}}

\def\R{{\cal R}}
\def\U{{\cal U}}

\def\Re{{\rm Re}}

\def\tr{{\rm tr}}
\def\det{{\rm det}}
\def\half{ {1\over 2}}
\def\quart{{1 \over 4}}

\def\p{\partial}

\def\ep{\varepsilon}

\def\sub{0}

\def\no{\nonumber}

\begin{document}
\starttext
\baselineskip=16pt
\setcounter{footnote}{0}

\begin{flushright}
UCLA/06/TEP/03 \\
01 March 2006
\end{flushright}

\bigskip

\begin{center}
{\Large\bf Interface Yang-Mills, Supersymmetry, and Janus}

\vskip .7in 

{\large  Eric D'Hoker,  John Estes and  Michael Gutperle}

\vskip .2in

 \sl Department of Physics and Astronomy \\
\sl University of California, Los Angeles, CA 90095, USA

\end{center}

\vskip .5in

\begin{abstract}

We consider theories consisting of a planar interface with $\N=4$ 
super-Yang-Mills on either side and varying gauge coupling across 
the interface.  The interface does not carry any independent degrees 
of freedom, but is allowed to support local gauge invariant  
operators, included with independent interface couplings.
In general, both conformal symmetry and supersymmetry will be broken, 
but  for special arrangements of the interface couplings, 
these symmetries may be restored. We provide a systematic classification
of all allowed interface supersymmetries. We find new theories preserving 
eight and four  Poincar\'e supersymmetries, which get extended to
sixteen and eight supersymmetries in the conformal limit,
respectively with $SU(2) \times SU(2)$, $SO(2) \times SU(2)$ 
internal symmetry.  The Lagrangians for these theories are explicitly constructed.
We also recover the theory with two Poincar\'e supersymmetries and $SU(3)$ 
internal symmetry proposed earlier as a candidate CFT dual to super Janus. 
Since our new interface theories have only operators from the supergravity 
multiplet turned on, dual supergravity solutions are expected to exist. 
We speculate on the possible relation between the interface theory
with maximal supersymmetry and the near-horizon limit of the D3-D5 system.

\end{abstract}

\vfill\eject

\baselineskip=16pt
\setcounter{equation}{0}
\setcounter{footnote}{0}

\newpage

\section{Introduction}
\setcounter{equation}{0}

The Janus solutions of Type IIB supergravity, found in \cite{Bak:2003jk}, 
form a continuous family of deformations of $AdS_5 \times S_5$, in which 
the dilaton is allowed to vary along a single space direction, but the 
anti-symmetric 2-form field $B_{(2)}$ is vanishes. The Janus solutions 
are smooth and singularity-free. They may be simply generalized to include a 
varying axion along with a varying dilaton, and admit an analytic representation 
in terms  of elliptic functions \cite{edjemg}. They are invariant under 
$SO(2,3) \times SO(6)$, but break  all supersymmetries. Remarkably, 
despite their lack of supersymmetry, the solutions are classically 
stable against all small and a certain class of large perturbations 
\cite{Bak:2003jk,Freedman:2003ax,Celi:2004st}.

\smallskip

The Janus solutions offer interesting candidates for further exploring the 
AdS/CFT correspondence \cite{Maldacena:1997re, Gubser:1998bc,
Witten:1998qj} (for reviews, see \cite{D'Hoker:2002aw,
Aharony:1999ti}). Some aspects of the holographic properties of the
Janus solution were discussed in
\cite{Bak:2003jk,Freedman:2003ax,Papadimitriou:2004rz}.  The boundary
of a  
Janus solution consists of two halves of Minkowski space-time which are joined 
along an interface. The dilaton takes two different asymptotic  values on each 
of these boundary components.  The CFT dual, proposed in 
\cite{Bak:2003jk, Clark:2004sb}, consists of a 3+1-dimensional gauge 
theory with a 2+1-dimensional planar interface.
The gauge theory on each side of the planar interface is $\N=4$ super Yang-Mills
and the gauge coupling varies discontinuously across the interface. 
The two values of the gauge coupling correspond to the two asymptotic values 
of the dilaton in the Janus solution.  
The non-constancy of the gauge coupling allows the action to include 
{\sl interface operators}, whose support is limited to the interface. 
These operators involve only $\N=4$ super-Yang-Mills fields, 
and the interface carries no independent degrees of freedom.  This is to be
contrasted with defect or boundary 
conformal field theories \cite{Cardy:1984bb,Cardy:1991tv,McAvity:1995zd},
where often  new degrees of freedom are localized on the defect.

\smallskip

At present, it is unclear whether the Janus solution arises as the near-horizon
limit of any interesting brane configuration. The absence of independent
degrees of freedom on the interface seems to preclude the presence of
open strings between the D3 branes of the original undeformed 
$AdS_5 \times S_5$ and the interface. Or, if open strings did arise,
the question becomes how they could have decoupled in the supergravity limit.

\smallskip

The absence of any degree of supersymmetry in the Janus solution 
makes it difficult to investigate such brane candidates directly. 
Therefore, one is interested in generalizing the original non-supersymmetric 
Janus to new solutions of Type IIB supergravity  
which do have non-trivial supersymmetry. Part of this goal was 
achieved in \cite{Clark:2004sb} where a supersymmetric Janus solution
was found in a five dimensional gauged supergravity theory with $N=2$
supersymmetry. In \cite{edjemg} supersymmetric Janus solutions of
ten dimensional Type IIB supergravity were 
found with $\N=1$ interface supersymmetry and $SU(3)$ internal symmetry. 
Clearly, however, it would be advantageous to understand not just 
one or a few such solutions, but rather the full space 
of supersymmetric Janus-like solutions. Progress in this direction can be 
made by analyzing the degree of supersymmetry in the dual CFT.

\smallskip

In this paper, we shall pose and solve the following problem.
Consider a gauge theory in 3+1 dimensions  with a 2+1-dimensional
planar interface,  with $\N=4$ super Yang-Mills on each side of
the interface, and no independent  degrees of freedom on the interface.
The action of the theory  allows for local gauge invariant interface
operators consistent with interface conformal invariance, to be included with 
arbitrary {\sl interface couplings}. The problem is to classify completely, 
as a function of the interface couplings, 
the degree of supersymmetry and conformal invariance of the theory. Our
investigation is close in spirit to \cite{Eto:2005mx,Eto:2005cp,Oda:1999az}.

\smallskip

A gauge coupling which is strictly discontinuous across the interface leads to
a number of technical subtleties and complications, such as  the possibility
of discontinuous canonical fields and the appearance of the square of Dirac  
$\delta$-functions \cite{Clark:2004sb}. Therefore, the problem posed in the 
preceding paragraph will be solved in two stages. In the first stage, the
gauge coupling will be allowed to vary smoothly across the  interface, and the
problem    
of determining the degree of Poincar\'e supersymmetry in the presence of this
smoothly varying gauge coupling will be solved. The supersymmetry
transformation rules in the presence of the interface will require certain
modifications from their standard form in $\N=4$ super Yang-Mills.  
In the second stage, the existence of the limit of this smoothly
varying gauge coupling to a discontinuous jump will be investigated.
When this limit exists,  interface conformal invariance will be recovered,
and  Poincar\'e supersymmetry will be enhanced to interface
superconformal symmetry.

\smallskip

The results obtained in this paper may be summarized as follows.
The conditions for Poincar\'e supersymmetry in the presence of a
smoothly varying gauge coupling are reduced to a complicated looking set 
of 7 algebraic matrix equations involving the gauge coupling, the 
interface couplings,  and the supersymmetry parameters. Remarkably,
these equations may be drastically simplified and their solutions
classified systematically. The existence of the theory \cite{Clark:2004sb}
with  2 real Poincar\'e supercharges, $SU(3)$ global symmetry, and interface conformal invariance is confirmed.
New solutions with 16 and 8 conformal supersymmetries are discovered, with respectively
$SU(2) \times SU(2)$ and $SO(2) \times SU(2)$ symmetry.

\smallskip

The rest of the paper is organized as follows.  In section 2, we define the general 
interface $\N=4$ Yang-Mills theory, list its allowed interface terms and 
couplings, and derive its global symmetries.  In section 3, we derive a set 
of algebraic equations for the existence of at least one interface
supersymmetry, 
which is solved in section 4. This result is used in section 5 to produce a
systematic classification of allowed Poincar\'e interface supersymmetries.
The existence of the conformal limit is discussed in section 6.
In section 7, a comparison is made between the new theory with 
16 interface supersymmetries and $SU(2)\times SU(2)$ internal symmetry, 
and the near-horizon limit of the  D3-D5 system.

\newpage

\section{$\N=4$ super Yang-Mills with an interface}
\setcounter{equation}{0}

In this section, we obtain a systematic generalization of  the interface 
Yang-Mills theories, 
proposed in \cite{Bak:2003jk, Clark:2004sb} as duals to the Janus solution. 
The goal of the generalization is to achieve a complete classification of the 
possible supersymmetries supported  by such interface theories. The 
generalization consists of 3+1-dimensional gauge theory, with a 2+1-dimensional
planar interface, $\N=4$ super Yang-Mills on each side of the interface,
no independent degrees of freedom on the interface, and local
gauge invariant interface operators included in the action with arbitrary
interface couplings. Since we are primarily interested in conformal 
theories, we shall assume throughout that no dimensionful  couplings 
enter into the action. This requirement also ensures renormalizability. 

\smallskip

We start with a summary of standard $\N=4$ super Yang-Mills 
\cite{Brink:1976bc, Gliozzi:1976qd}.  The theory contains
a gauge field $A_\mu$, four Weyl fermions $\psi ^a$, and six real scalars $\phi^i$, which transform under $SU(4)$ in the $\bf 1$, $\bf 4$, and $\bf 6$ 
representations respectively, and under the $SU(N)$ gauge 
group in the adjoint representation.\footnote{The fields $A_\mu$, $\psi ^a$
and $\phi^i$ take values in the gauge algebra. Internal labels will be denoted
by Latin indices $a=1,\cdots, 4$, and $i=1,\cdots ,6$. Each Weyl spinor $\psi ^a$ 
will be expressed as a 4-component Dirac spinor whose right chirality component vanishes, and the 4 Weyl spinors $\psi^a$ will be grouped in a quadruplet
for which we use the matrix notation $\psi$.}  
The Lagrangian is given by
\bea
\L_\sub &=& 
 - {1 \over 4 g^2} \tr \left ( F^{\mu \nu } F_{\mu \nu } \right ) 
- {1 \over 2 g^2} \tr (D^\mu \phi^i D_\mu \phi^i )
+ { 1 \over 4 g^2} \tr ( [\phi^i, \phi^j ]  [\phi ^i, \phi ^j ] )
\no \\ &&
- {i \over 2 g^2} \tr \left ( \bar \psi \gamma ^\mu D_\mu \psi \right )
+ {i \over 2 g^2} \tr \left ( D_\mu \bar \psi \gamma ^\mu \psi \right )
\no \\ &&
+ {1 \over 2 g^2} \tr \left ( \psi ^t \C  \rho^i [\phi ^i, \psi] 
+ \psi^\dagger \C (\rho^i)^*  [\phi ^i, \psi ^* ] \right )
\eea
The trace is over the gauge algebra only. The $\gamma ^\mu$ are the 
Dirac matrices, and $\C$ is the associated charge conjugation matrix  
defined by $(\gamma ^\mu)^t = - \C \gamma^\mu \C^{-1}$.
The $4 \times 4$ matrices $\rho^i$ are  Clebsch-Gordan coefficients for the $SU(4)$ tensor product decomposition ${\bf 4} \otimes {\bf 4} \to {\bf 6}$. 
Along with other useful $SU(4)$ matrices, the $\rho ^i$ are presented in  
Appendix A, where they are obtained from  the Clifford algebra in  6 Euclidean dimensions. The gauge coupling $g$ is a constant and the $CP$-violating 
$\tr(F \tilde F)$ term has been omitted. 

\smallskip

The Lagrangian $\L_\sub$ is invariant under $SU(4)$ R-symmetry, 
conformal $SO(2,4)$, and $\N=4$ Poincar\'e supersymmetry, which is 
enhanced to $\N=4$ conformal supersymmetry.  These symmetries combine 
into a simple supergroup $PSU(2,2|4)$.

\smallskip

A planar interface, whose position is described by the vanishing of a single
space coordinate, $x^\pi$, may be introduced by making the coupling
discontinuous across $x^\pi =0$, and adding local gauge invariant interface
operators of dimension 3 with arbitrary interface couplings, but no 
independent degrees of freedom on the interface.   In the context of the 
AdS/CFT correspondence, such an interface will not allow strings stretching 
between the interface and  D3-branes. With this set-up, the interface will 
preserve gauge invariance, as well as the $SO(2,3)$ conformal invariance 
which leaves the plane  $x^\pi=0$ invariant. For general interface 
couplings, $SU(4)$ and supersymmetry will be broken. 

\smallskip

It was remarked already in \cite{Clark:2004sb}, however, that a strictly 
discontinuous gauge coupling may introduce technical complications, 
such as discontinuous canonical fields and squares of Dirac $\delta$-functions 
at $x^\pi =0$.
Therefore, following \cite{Clark:2004sb}, we introduce a gauge coupling
function $g(x^\pi)$ which varies smoothly across the interface, such
as represented schematically in Fig 1. In the presence of a smooth 
gauge coupling, no technical subtleties will arise, and the interface 
theory will be well-defined. It will be natural, not to localize interface
operators strictly at $x^\pi=0$, but instead to introduce them as
4-dimensional operators whose coupling is proportional to the derivative
of the gauge coupling $\p_\pi g$. Of course, any smooth, non-constant, 
$g(x^\pi)$ has at least
one length scale and therefore will break $SO(2,3)$ conformal invariance 
further to 2+1-dimensional Poincar\'e invariance. Conformal invariance may
then be recovered if the limit in which $g(x^\pi)$ tends to a step function
actually exists and makes sense.  It is interesting to note that our supersymmetry
analysis will hold for a general gauge coupling varying along $x^\pi$.

\begin{figure}[tbph]
\begin{center}
\epsfxsize=4.0in
\epsfysize=2in
\epsffile{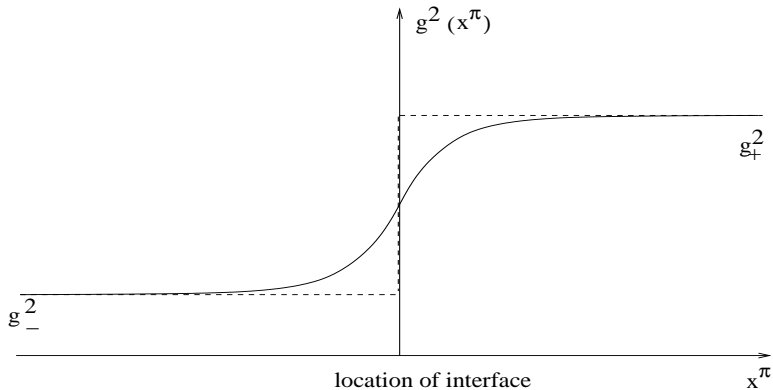}
\label{figure1}
\caption{The discontinuous gauge coupling is replaced by a smoothly varying function.}
\end{center}
\end{figure}

It remains to give a complete description of the local gauge invariant 
interface operators and their couplings. We introduce a length scale 
only through the gauge coupling function $g(x^\pi)$, so that the interface 
operators will always be multiplied by at least one power of $\p_\pi g$. 
This will guarantee that the interface Lagrangian will not contribute to the  
dynamics in the bulk.  We write the full Lagrangian as follows,
\bea
\label{Ltotal}
\L = \L_\sub + \L_I
\eea
where $\L_I$ stands for the interface Lagrangian. We now list the possible 
interface operators in $\L_I$, consistent with gauge invariance,
2+1-dimensional Poincar\'e invariance, and of dimension no higher than 3. 
It is useful to organize the contributions as follows,
\bea
\label{Linterface}
\L _\psi & = & 
{(\partial_\pi g) \over g^3} \tr \left (
y_1 \bar \psi \gamma ^\pi \psi 
+ {i \over 4} y_2^{ij} \bar \psi \gamma^\pi \rho^{ij} \psi 
- {i \over 2} y_3^{ijk} \left (\psi^t  \C \rho^{ijk} \psi 
+ \psi^\dagger \C (\rho^{ijk})^* \psi^* \right ) \right)
\no \\
\L _\phi & = &
 {(\partial_\pi g) \over 2g^3} \tr \left (
 z_1^{ij} \p_\pi (\phi ^i \phi ^j)
 +  2 z_2^{ij} \phi ^{[i} D_\pi \phi ^{j]} 
 - i z_3^{ijk} \phi^i [\phi^j,\phi^k] 
   \right)
\no \\
\L_{\phi^2} & = &
{(\partial_\pi g)^2 \over 2g^4} z_4^{ij} \, \tr \left ( \phi^i \phi^j \right )
\no \\
\L _{{\rm CS}} & = &
{(\p_\pi g) \over g^3} z_{CS} \, \ep ^{\pi \mu \nu \rho} 
\tr \left ( A_\mu \p_\nu A_\rho + { 2 \over 3} i A_\mu A_\nu A_\rho \right )
\eea
Here, the interface couplings are all real functions of $x^\pi$.
The couplings $y_2^{ij}$, $y_3 ^{ijk}$, $z_2^{ij}$,  and  $z_3 ^{ijk}$
are totally anti-symmetric in their indices, while $z_1 ^{ij}$, and $z_4 ^{ij}$ 
are totally symmetric. 
Overall factors of 2 and $g$ have been introduced for later convenience.
The following abbreviated notations will be useful,
\bea
\label{Y2Y3}
Y_2 \equiv  y_2 ^{ij} \rho ^{ij}
& \hskip 1in & 
Y_3 \equiv  - y_3 ^{ijk} \left ( \rho ^{ijk} \right )^*
\no \\
Z_2 \equiv  z_2 ^{ij} \rho ^{ij}
& \hskip 1in & 
Z_3 \equiv  - z_3 ^{ijk} \left ( \rho ^{ijk} \right )^*
\eea
By construction, and using (\ref{conjugation}) of the Appendix, we have 
$Y_2 ^\dagger =- Y_2$, $Z_2 ^\dagger = - Z_2$,
$Y_3^t = Y_3$, $Z_3 ^t = Z_3$, and $\tr (Y_2) = \tr (Z_2)=0$.
Under $SU(4)$, the interface operators and couplings transform
in the following representations,
\bea
\label{interface couplings}
&&      y_1: {\bf 1}        \hskip .98in
z_1: {\bf 1} \oplus {\bf 20} ^\prime
    \hskip .6in 
    z_4: {\bf 1} \oplus {\bf 20} ^\prime 
    \no \\
&&      y_2: {\bf 15}  \hskip .9in
    z_2: {\bf 15}  \hskip .9in
     z_{CS}: {\bf 1}
    \no \\
&&      y_3: {\bf 10} \oplus \overline {\bf 10}   \hskip .5in
       z_3: {\bf 10} \oplus {\bf \overline{10}} 
\eea
In the context of AdS/CFT, the $\bf 10$, $\bf \overline{10}$, $\bf 15$, and
    $\bf 20'$, correspond to supergravity fields, while the $\bf 1$
    corresponds to higher string modes. We do not expect the latter to be  
relevant in supergravity solutions, but shall include them here for
    completeness. 
Since we omitted the $CP$-violating term $\tr (F \tilde F)$ from $\L_\sub$,
we shall omit its $CP$-violating interface counterpart $\L_{CS}$. 

\smallskip

The final interface contributions to the Lagrangian of interest to us will be,
\bea
\label{LI}
\L _I = \L_\psi + \L_\phi + \L_{\phi ^2}
\eea
It is clear that all interface terms in $\L_\psi$, $\L_\phi$ correspond to 
local gauge invariant, dimension 3 operators whose contribution will be
invariant under $SO(2,3)$ conformal transformations in the limit where
$g$ tends to a step function.

\smallskip

It remains to establish that also the converse is true, namely that we have 
indeed retained all allowed operators. Combining gauge invariance
and 2+1-dimensional Poincar\'e invariance with the requirement that the 
dimension of the interface operators can be at most 3, it is clear that no 
operators of dimension 1 can occur (since $\tr (\phi ^i)=0$), and that the only dimension 2 operator is $\L_{\phi^2}$. Dimension 3 operators can then be (1)  bilinears in the Fermi field, (2) trilinears in the  scalar field or (3) bilinears in 
the scalar field with one extra derivative acting. $\L_\phi$ contains all possible 
terms in groups (2) and (3). The term $\tr (\phi ^i \{ \phi ^j, \phi ^k \})$,
which is totally symmetric in $i,j,k$, and which is a candidate for group (2),
actually vanishes since the field $\phi^i$ takes values in the adjoint 
representation of  the  gauge algebra of $SU(N)$, (which is real) so that 
$\tr (\phi ^i \{ \phi ^j, \phi ^k \})=0$. Finally, two types of fermion bilinears are allowed
by $SO(1,2)$ Lorentz invariance, and the fact that $\psi$ is a Weyl spinor,
namely $\tr(\bar \psi \gamma ^\pi M_1 \psi)$ and $\tr(\psi ^t \C M_2 \psi)$ plus its 
complex conjugate. Here, $M_1$ and $M_2$ are $4\times 4$ Clebsch-Gordan matrices of $SU(4)$, which in view of $CPT$-invariance and transposition symmetry of the bilinears must satisfy $M_1 ^\dagger = M_1$ and $M_2 ^t=M_2$.
The unique solutions of these requirements are the ones listed in $\L_\psi$. 

\smallskip

The interface couplings $y_2^{ij}$ and $z_2 ^{ij}$ may be viewed
as external $SU(4)\sim SO(6)$ gauge fields which are localized at
the interface, while $y_1$ may be viewed as an external $U(1)$ gauge
field. Changing the fields $\phi ^i$ and $\psi$ by a local $SU(4)$
transformation which depends only on $x^\pi$ will have the effect of
making a gauge transformation on $y_2 ^{ij}$, $z_2 ^{ij}$, and $z_4 ^{ij}$.
These transformations are free of anomalies and their existence
shows that a given physical theory may be represented via 
different sets of interface couplings, as long as they are related 
by such an $SU(4)$ gauge transformation.

\smallskip

In the formulation of interface Lagrangians given above, it is straightforward
to recover the two interface CFTs proposed in \cite{Clark:2004sb} as duals to the Janus solution.
The first Lagrangian is obtained by choosing $z_1^{ij} = \delta^{ij}$ and all other parameters to be zero, while the second one is recovered by also choosing 
$z_4^{ij} = - \delta^{ij}$ and then redefining the scalar fields by 
$\phi^i \to \phi^i/g$.  Interestingly, the $(\p_\pi g)^2$ term now drops 
out for the redefined scalar fields, so that the limit to a step function 
gauge coupling is well defined.  We shall list the general conditions for 
the existence of the conformal limit later.

\newpage

\section{Supersymmetry in the presence of an interface} 
\setcounter{equation}{0}

In this section, we discuss the fate of supersymmetry in the interface
Yang-Mills theory governed by the Lagrangian $\L$ of (\ref{Ltotal}) whose
interface operators and couplings are given in (\ref{Linterface}) and (\ref{LI}).
We reduce the conditions for the existence of supersymmetry to a set
of algebraic equations involving the gauge coupling, the interface couplings
and the supersymmetry parameter. These equations will be solved in 
subsequent sections.

\smallskip

The supersymmetry transformations for the Lagrangian $\L_\sub$
are as follows \cite{Brink:1976bc, Gliozzi:1976qd},\footnote{The complex 
conjugation matrix $B$ is defined to obey the relation 
$\B \gamma _\mu \B ^{-1} = (\gamma _\mu )^*$; 
it may be chosen so 
that $\B ^*=\B $, $\B ^2=I$, and is given  by $\B = \C \gamma ^0$.}
\bea
\label{susy1}
\delta_\sub A_\mu  
&=& i \bar \psi \gamma_\mu \zeta - i \bar \zeta \gamma_\mu \psi 
\no \\ 
\delta_\sub \phi^i 
&=& i \zeta^t  \C \rho ^i \psi + i \bar\psi \B (\rho ^i)^*  \zeta^* 
\no \\
\delta_\sub \psi 
&=& + \half F_{\mu \nu} \gamma^{\mu \nu} \zeta 
+ (D_\mu \phi^i) \gamma^\mu \B  (\rho ^i)^*  \zeta^* 
- {i \over 2} [\phi^i,\phi^j] \rho^{ij} \zeta 
\eea
It will be useful to have the transformations of the following fields as well,
\bea
\label{susy2}
\delta _\sub F_{\mu \nu} 
& = &
i D_\mu \bar \psi \gamma _\nu \zeta - i \bar \zeta \gamma _\nu D_\mu \psi
- i D_\nu \bar \psi \gamma _\mu \zeta + i \bar \zeta \gamma _\mu D_\nu \psi
\no \\
\delta _\sub (D_\mu \phi^i) 
& = &
i \zeta ^t \C \rho ^i  D_\mu \psi 
+ i \zeta ^\dagger \C (\rho ^i)^* D_\mu \psi ^*
+ [ \bar \psi \gamma _\mu \zeta  - \bar \zeta \gamma _\mu \psi , \phi ^i]
\no \\
\delta_\sub \bar \psi 
&=& - \half  \bar\zeta \gamma^{\mu \nu}  F_{\mu \nu} 
- (D_\mu \phi^i)  \zeta^t  \C \rho ^i   \gamma^\mu 
+ {i \over 2} [\phi^i,\phi^j] \bar \zeta \rho^{ij}
\eea
Under the transformation $\delta _\sub$, with space-time dependent 
supersymmetry parameter $\zeta $, the Lagrangian transforms 
as follows,\footnote{The customary Majorana notation $(\p_\kappa \bar \zeta) S^\kappa$  includes the contribution from $\zeta$ and $\zeta ^*$.}
\bea
\label{N=4trans}
\delta _\sub \L_\sub = {1 \over g^2} \left (
\p _\kappa X^\kappa - (\p_\kappa \bar \zeta ) S^\kappa \right )
\eea
where
\bea
X^\kappa & = & 
{i \over 4} \tr \left( 
F_{\mu \nu} \bar\psi \gamma^{\mu \nu} \gamma^\kappa \zeta 
- 2 D_\mu \phi^i \bar\psi \gamma  ^\mu  \gamma ^\kappa  \B (\rho^i)^* \zeta^*  
-i [\phi ^i, \phi ^j] \bar \psi   \gamma ^\kappa  \rho^{ij} \zeta \right) + \mbox{c.c.}
\\
(\partial_\pi \bar\zeta) S^\kappa  &=& 
 {i \over 2} \tr \bigg(
F_{\mu \nu } \bar \psi \gamma^\kappa \gamma^{\mu \nu} \p_\kappa \zeta  
+ 2  D_\mu \phi ^i \bar \psi \gamma^\kappa \gamma^\mu \B (\rho^i)^* \p_\kappa \zeta^* 
- i [\phi ^i, \phi ^j] \bar \psi \gamma^\kappa \rho^{ij} \p_\kappa \zeta \bigg)
+ \mbox{c.c.}
\no
\eea
For constant $g$, it is clear from the transformation (\ref{N=4trans}), that 
the Lagrangian $\L_\sub$ is invariant under the global symmetry generated 
by $\delta _\sub$ with constant $\zeta$, for which $S^\kappa$ is the conserved supercurrent. Indeed, for constant $g$ and $\zeta$, the Lagrangian 
changes by a total derivative $\p_\kappa (g^{-2} X^\kappa)$, signaling that
$\delta_\sub$ is a symmetry.

\subsection{Conditions for interface supersymmetry}

For non-constant $g$, (such as when $g$ is a function of $x^\pi$)
the Lagrangian $\L_\sub$ no longer changes by a total derivative
under the transformation $\delta _\sub$ with constant $\zeta$,
but instead we have now the following transformation equation, 
\bea
\delta _\sub \L_\sub = \p_\kappa \left (
{1 \over g^2} X^\kappa \right ) 
- {1 \over g^2} (\p_\kappa \bar \zeta ) S^\kappa 
+ 2 {\p_\kappa g \over g^3} X^\kappa
\eea
As a result, supersymmetry will be broken. For an interface described 
by a gauge coupling $g(x^\pi)$, the obstructing term is proportional 
to $\p_\pi g$ and is thus localized on the interface. Therefore, it is natural
to modify the Lagrangian by including the interface operators of
(\ref{Linterface}) and (\ref{LI}),  to modify the supersymmetry
transformations by interface terms $\delta _I$ proportional to $\p_\pi g$,
\bea
\label{delta}
\delta = \delta _\sub +  \delta _I
\eea
and to let the supersymmetry parameter depend on $x^\pi$.
Restricting the form of $\delta _I$ by 2+1-dimensional Poincar\'e 
invariance, gauge invariance and the requirement that no extra 
dimensionful parameters should enter, we find that,
\bea
\delta _I \phi^i = \delta _I A_\mu = 0
\no \\
\delta _I \psi = (\p_\pi g) \chi ^i \phi ^i
\eea
The spinors\footnote{$\chi^i$ is a Weyl spinor, in the ${\bf 4}$ of $SU(4)$, and is a singlet under the $SU(N)$ gauge symmetry.} $\chi^i$ have scaling dimension $-1/2$, and are as of yet 
undetermined functions of $x^\pi$.  The scaling dimension is due to a factor of $\zeta$,
which will be shown later (\ref{summary}), so that $\chi^i$ does not introduce any
dimensionful parameters.  The requirement of supersymmetry 
of the full Lagrangian $\L$ of (\ref{Ltotal}) under the full transformations
$\delta$ of (\ref{delta}) is that the full variation,
\bea
\delta \L = \delta _\sub \L_\sub + \delta _\sub \L_I + \delta _I \L_\sub + \delta _I \L_I
\eea
be a total derivative. Since $g$ and $\zeta$ are smooth functions of only the single 
space-coordinate $x^\pi$, this condition reduces to
\bea
\label{mastereq}
- {1 \over g^2} (\p_\pi \bar \zeta ) S^\pi + 2 {\p_\pi g \over g^3} X^\pi
+ \delta _\sub \L_I + \delta _I \L_\sub + \delta _I \L_I \sim 0
\eea
up to a total derivative. The calculations of $\delta _\sub \L_I$, $ \delta _I \L_\sub$,
and $ \delta _I \L_I$ are straightforward, but the resulting expressions are 
lengthy and will not be presented here.

\smallskip

Besides total derivative terms, the equation (\ref{mastereq}) involves 
different functionally independent combinations of the canonical fields 
$\phi ^i$, $\psi$, $\bar \psi$, and $A_\mu$, which must vanish independently. 
Half of these terms involve $\psi$, the other half involve $\bar \psi$, which are complex conjugates of one another. 
It suffices to enforce the vanishing of the terms in $\bar \psi$, as the ensuing
equations will, by complex conjugation, also imply the vanishing of the
terms in $\psi$. The independent field combinations are
$\phi^i \bar \psi $, $\phi ^i D_\pi \bar \psi$, $F_{\mu \nu}\bar \psi $, 
$(D_\mu \phi ^i) \bar \psi $, and $[\phi^j, \phi^k] \bar \psi $.\footnote{Note that
$D_\pi(\phi \bar\psi)$ will not yield a total derivative in the action so that 
$\phi ^i D_\pi \bar \psi$ and $(D_\mu \phi ^i) \bar \psi $ are functionally
independent.}  Their 
coefficients  must independently vanish, and yield the following 
equations;\footnote{Throughout, we continue to use the notations of 
(\ref{Y2Y3}) for $Y_2$ and $Y_3$.}

\smallskip

\noindent
$\bullet$  The coefficients of  $\phi ^i \bar \psi$ are,
\bea
\label{reduce1}
0 =
(-i y_1 +  \quart Y_2 -1 ) \gamma^\pi \chi^i
-   \B Y_3 (\chi^i)^*
+ z_4^{ij}  \B (\rho^j)^* \zeta^*
+ (z_1^{ij} + z_2^{ij})  \B (\rho^j)^* {\partial_\pi \zeta^* \over \partial_\pi g}
\eea

\smallskip

\noindent
$\bullet$ The coefficients of $\phi ^i D_\pi \bar \psi$ are,
\bea
\label{reduce2}
0 =
(z_1^{ij} + z_2^{ij}  ) \B (\rho^j)^* \zeta^*
+  \gamma^\pi \chi^i
\eea

\smallskip

\noindent
$\bullet$  The coefficients of $ F_{\mu \nu}\bar\psi $ are,
\bea
\label{reduce3}
0 = 
 \gamma^{\mu \nu} \gamma^\pi \zeta
-  \gamma^\pi \gamma^{\mu \nu} {\partial_\pi \zeta \over \partial_\pi g}
-i y_1 \gamma ^\pi \gamma ^{\mu \nu} \zeta
+ \quart  \gamma^\pi \gamma^{\mu \nu} Y_2 \zeta
-  \gamma^{\mu \nu} \B Y_3 \zeta^* 
\eea

\smallskip

\noindent
$\bullet$  The coefficients of $\bar\psi (2 D_\mu \phi^i)$ are,
\bea
\label{reduce4}
0 &=&
-  \gamma^\mu \gamma^\pi \B (\rho^i)^* \zeta^*
-  \gamma^\pi \gamma^\mu \B (\rho^i)^* {\partial_\pi \zeta^* \over \partial_\pi g}
-i y_1 \gamma^\pi \gamma^\mu \B (\rho^i)^* \zeta^* 
+ \quart  \gamma^\pi \gamma^\mu \B Y_2 (\rho^i)^* \zeta^*
\no\\ &&
+  \gamma^\mu Y_3 \rho^i \zeta 
- z_2^{ij}  \B (\rho^j)^* \zeta^* \delta^{\mu \pi}
+ z_1^{ij}  \B (\rho^j)^* \zeta^* \delta^{\mu \pi}
-  \gamma^\mu \chi^i  
+  \gamma^\pi \chi^i \delta^{\mu \pi} 
\eea 

\smallskip

\noindent
$\bullet$  The coefficients of $ i[\phi^i,\phi^j] \bar\psi$ are,
\bea
\label{reduce5}
0 &=&
- \gamma^\pi \rho^{ij} \zeta
+ \gamma^\pi \rho^{ij} {\partial_\pi \zeta \over \partial_\pi g}
+ i y_1  \gamma^\pi \rho^{ij} \zeta 
- \quart \gamma^\pi Y_2 \rho^{ij} \zeta
+ \B Y_3 (\rho^{ij})^* \zeta ^*
\no \\ &&
- 3 z_3^{ijk}  \B (\rho^k)^* \zeta^*
+ 2 z_2^{ij}  \gamma_\pi \zeta
+ \B (\rho^i)^* (\chi^j)^*
- \B (\rho^j)^* (\chi^i)^*
\eea
In (\ref{reduce1}), we have omitted a common factor of $i (\partial_\pi g)^2$, 
while in (\ref{reduce2}-\ref{reduce5}), we have omitted a common factor of 
$i (\p_\pi g)/2$.
Finally, all terms with no $\partial_\pi g$ dependence are total derivatives 
and do not contribute to the variation of the action.

\subsection{Simplifying the Conditions for Interface Supersymmetry}

We shall now simplify and to some extent decouple the equations 
for interface supersymmetry, obtained in (\ref{reduce1}-\ref{reduce5}). 
The result is the following group of 7 equations,
\bea
\label{summary}
(1) & \hskip .1in & 
    \zeta =  Y_3 \gamma^\pi \B \zeta^*
\no \\
(2) & \hskip .1in &
Y_3 \rho^i \zeta = 
-  (\rho^i)^* \gamma^\pi \B \zeta^* 
+ z_1^{ij}  (\rho^j)^* \gamma^\pi \B \zeta^*
\no \\
(3) & \hskip .1in &
     - \rho^{ij} \zeta + y_1 i \rho^{ij} \zeta
     + \rho^{ij} {\p_\pi \zeta \over \p_\pi g} 
     - \quart Y_2 \rho^{ij} \zeta 
     +  Y_3 (\rho^{ij})^* \gamma^\pi \B \zeta^*  
     - 3 z_3^{ijk}   (\rho^k)^* \gamma^\pi \B \zeta^* 
\no\\&& \hskip .4in
     + 2 z_2^{ij}  \zeta 
     - (\rho^{i})^* (z_1^{jk} + z_2^{jk}) \rho^k \zeta
     + (\rho^{j})^* (z_1^{ik} + z_2^{ik}) \rho^k \zeta = 0
\no \\
(4) & \hskip .1in &
    {\p_\pi \zeta \over \p_\pi g}  = - i y_1 \zeta + \quart Y_2 \zeta
\no \\
(5) & \hskip .1in &
    \rho^i {\p_\pi \zeta \over \p_\pi g} = 
    i y_1 \rho^i \zeta + \quart Y_2^* \rho^i \zeta - z_2^{ij} \rho^j \zeta
\no\\
(6) & \hskip .1in &
     \chi^i = -  (z_1^{ij} + z_2^{ij}) (\rho^j)^* \gamma^\pi \B \zeta^*
\no\\
(7) & \hskip .1in &
     (-i y_1 + \quart Y_2 -1 )(z_1^{ij} + z_2^{ij}) (\rho^j)^* \gamma^\pi \B \zeta^*  
    - Y_3 (z_1^{ij} + z_2^{ij}) \rho^j \zeta 
    \no\\ && \hskip .4in
    - z_4^{ij} (\rho^j)^* \gamma^\pi \B \zeta^* 
    - (z_1^{ij} + z_2^{ij}) (\rho^j)^*  {\p_\pi \gamma^\pi \B \zeta ^* \over \p_\pi g} = 0
\eea
These equations were obtained as follows. Multiplying (\ref{reduce2}) by
$\gamma ^\pi$ yields equation (6), which now gives $\chi^i$ in terms of the 
other variables.
Equation (6) may be used to eliminate $\chi^i$ from all other equations;
doing so in  (\ref{reduce1}) yields equation (7), while doing so in (\ref{reduce5}) 
yields equation (3). 

\smallskip

To simplify (\ref{reduce3}), we use the fact that, as $\mu$  and $\nu$
run through their possible values $0,1,2,3$, the Clifford generators
$\gamma ^{\mu \nu} \gamma ^\pi$ are generally linearly independent from 
$\gamma ^\pi \gamma ^{\mu \nu}$, so that their coefficients in (\ref{reduce3}) 
must vanish independently; this yields equations (1) and (4) respectively. 
Equivalently, one may consider the two distinct cases $\pi \in \{ \mu, \nu \}$
and $\pi \not \in \{ \mu, \nu \}$; equations (1) and (4) are then obtained
as the sum and difference of these two conditions.

\smallskip

Similarly, to simplify (\ref{reduce4}), we use the fact that, as $\mu$ 
run through $0,1,2,3$, the Clifford generators
$\gamma ^\mu \gamma ^\pi$ are generally linearly independent from 
$\gamma ^\pi \gamma ^\mu $, so that their coefficients in (\ref{reduce4}) 
must vanish independently. Using in these equations also (6) to eliminate 
$\chi^i$, yields (2) and (5) respectively, after some simplifications.
Equivalently, one may consider the two distinct cases $\pi=\mu$
and $\pi \not= \mu$; equations (2) and (5) are then obtained
as the sum and difference of these two conditions.

\subsection{Resolving the space-time spinor structure}

The preceding equations involve the spinors $\zeta$ and $\gamma ^\pi \B \zeta ^*$
with coefficients which act on the internal $SU(4)$ labels of the spinors,
but not on their 3+1-dimensional space-time spinor labels. Clearly,
it will be convenient to decompose the spinors $\zeta$ and $\gamma ^\pi \B \zeta ^*$
onto a basis in which $\gamma ^\pi \B$ is diagonal. This is achieved in this subsection, and the reduced equations are then derived. For definiteness, 
we choose $\B$ to obey the relations of footnote $\# 2$, namely 
$\B^* = \B^t = \B^{-1} =\B $, so that we have 
\bea
(\gamma ^\pi \B )^t = \gamma ^\pi \B
\hskip 1in
(\gamma ^\pi \B)^* = \B \gamma ^\pi = (\gamma ^\pi \B ) ^{-1}
\eea
As a result, the matrix $\gamma ^\pi \B $ is unitary, 
and has eigenvalues of unit modulus. Furthermore,  $\gamma ^\pi \B $ 
commutes with the 3+1-dimensional chirality matrix $\gamma$. 
Finally, $\det (\gamma ^\pi \B )=1$, while on the $+$ chirality subspace, 
we have $\det (\gamma ^\pi \B )_+=-1$.

\smallskip

Using the above facts about $\gamma ^\pi \B $, we conclude that, on the 
$+$ chirality subspace, $\gamma ^\pi \B$ has two  eigenvalues, 
$\lambda^2$ and $ - (\lambda^2) ^*$, with $\lambda \lambda ^* =1$. 
Since $\gamma ^\pi \B$ is unitary, it is diagonalizable, and we 
denote its eigenvectors by $s_\pm$,
\bea
\gamma ^\pi \B s _+ & = &  + \lambda^2  s_+ 
\no \\
\gamma ^\pi \B s _- & = &  - (\lambda^2) ^*  s_- 
\eea
Using $(\gamma ^\pi \B )^* = \B \gamma ^\pi$ and $\lambda \lambda ^* =1$, 
it follows that 
\bea
\gamma ^\pi \B s _+^* & = &  + \lambda^2  s_+^* 
\no \\
\gamma ^\pi \B s _-^* & = &  - (\lambda^2) ^*  s_- ^*
\eea
When the two eigenvalues are different ($\Re(\lambda^2 ) \not=0$), 
the eigenspaces are 1-dimensional, and we may choose $s_+ ^* = s_+$ 
and $s_- ^* = s_-$. When $\Re(\lambda^2 ) =0$, $\gamma ^\pi \B$ 
is proportional to the identity and the same choice is still valid.  
We now decompose $\zeta$ and $\p_\pi \zeta $ as follows,
\bea
\label{zetadec}
\zeta & = &  \lambda s_+ \otimes \xi_+ + i  \lambda ^* s_- \otimes \xi_-
\no \\
\p_\pi \zeta  & = & 
\lambda s_+ \otimes \p_\pi \xi_+ + i  \lambda ^* s_- \otimes \p_\pi \xi _- 
\eea
Applying the above rules for complex conjugation, we then derive,
\bea
\label{zetadecbar}
\gamma ^\pi \B \zeta^* & = & 
    \lambda s_+ \otimes \xi_+^*  + i \lambda^* s_- \otimes \xi_-^* 
\no \\
\gamma ^\pi \B \p_\pi \zeta^* & = & 
 \lambda s_+ \otimes \p _\pi \xi _+^*  
+ i \lambda^* s_- \otimes \p _\pi \xi _-^* 
\eea

\subsection{The reduced conditions for interface supersymmetry}

The equations of (\ref{summary}) may be decomposed onto the basis of
(\ref{zetadec}) and (\ref{zetadecbar}). 
It is manifest that both $\xi_+$ and $\p _\pi \xi_+$, and $\xi_-$ and $\p _\pi \xi _-$ 
satisfy the same equations. Dropping the $\pm$ subscripts on 
$\xi$ and $\p _\pi \xi$, we get 
\bea
\label{aleq}
(1') & \hskip .1in & 
    \xi = Y_3 \xi^*
\no\\
(2') & \hskip .1in &
Y_3 \rho^i \xi = - (\rho^i)^* \xi^* + z_1^{ij} (\rho^j)^* \xi^*
\no \\
(3') & \hskip .1in &
     \left ( \rho^{ij} Y_3 + Y_3 (\rho^{ij})^* 
     - (\rho^{i})^* Y_3^* (\rho^j)^* 
     + (\rho^{j})^* Y_3^* (\rho^i)^* \right ) \xi ^*
     - 3 z_3^{ijk}  (\rho^k)^* \xi^*
      \no\\&& \hskip .4in 
     + \quart \left ( \rho^{ij} Y_2 - Y_2 \rho^{ij} \right ) \xi 
     + 2 z_2^{ij}  \xi 
     - (\rho^{i})^* z_2^{jk} \rho^k \xi
     + (\rho^{j})^* z_2^{ik} \rho^k \xi
      = 0
\no \\
(4') & \hskip .1in &
    \p_\pi \xi  = (\p _\pi g) \left ( - i y_1 \xi + \quart Y_2 \xi \right )
\no \\
(5') & \hskip .1in &
2 i y_1 \rho^i \xi 
+ \quart \left( Y_2^* \rho^i - \rho^i Y_2 \right) \xi - z_2^{ij} \rho^j \xi =0
\no\\
(6\, ) & \hskip .1in &
     \chi^i = -  \left ( z_1^{ij} + z_2^{ij} \right ) (\rho^j)^* \gamma^\pi \B \zeta^*
\no\\
(7') & \hskip .1in &
\left ( z_4^{ij}  + (z_1^{ik} + z_2^{ik})(z_1 ^{jk} + z_2^{jk} ) \right ) (\rho^j)^* \xi^* =0 
\eea
We have used $(4')$ to eliminate $\p _\pi \xi$ from the equations 
leading to $(3')$, $(5')$, and (7'); used $(2')$ to eliminate 
$z_1^{ij}$ from the equation leading to $(3')$, and $Y_3 \rho ^j \xi$ from
the equation leading to (7'); used (5') to further simplify the 
equation that leads to (7'). Finally, (6) has been left undecomposed, as it 
merely gives an expression for $\chi^i$. For every $\xi$ satsifying these 
equations, we will have two real supersymmetries given by $\xi_+$ and $\xi_-$.

\smallskip

Given the assumption that a non-trivial supersymmetry survive
requires solutions with $\xi \not=0$. This in turn allows for a simple 
solution to equation (7')
\bea
\label{seven'}
(7') \qquad \qquad z_4^{ij} = - (z_1^{ik} + z_2^{ik})(z_1^{jk} + z_2^{jk})
\eea
Since the variable $z_4^{ij}$ does not appear in the remaining equations, 
we can take (\ref{seven'}) as the resulting value of $z_4^{ij}$.

\newpage

\section{Solving the Interface Supersymmetry Conditions}
\setcounter{equation}{0}

In this section, we shall solve the conditions for interface supersymmetry 
which, in the preceding section, were reduced to a set of algebraic equations (\ref{aleq}). To simplify the associated calculations, we shall use the covariance 
of the interface couplings, supersymmetry parameter, and equations under 
global and local $SU(4)$ transformations.

\subsection{Transformations under global $SU(4)$}

Let $U$ and $R$ be the representation matrices respectively in the 
${\bf 4}$ and ${\bf 6}$ of $SU(4)$, satisfying  $U^\dagger U=I$, $R^t R=I$. The fields and supersymmetry parameters transform as
\bea
\psi \to ~ \psi ' \hskip .1in = U \psi \hskip .2in & \hskip 1in & \zeta \to \zeta ' = U \zeta
\no \\
\phi ^i \to (\phi ')^i= R^{ii'} \phi ^{i'} & \hskip 1in & \xi \to \xi ' = U \xi
\eea
The $\rho$-matrices are invariant in the following sense,\footnote{This result 
may be derived by starting from the $SO(6)$ Dirac matrices $\gamma ^i_{(6)}$,
for which we have $\gamma ^i_{(6)} = R^{ii'} U \gamma ^{i'} _{(6)} U^\dagger$
and $U^* = \C_{(6)} U \C^{-1}_{(6)}$, and using the relation (\ref{rhogamma}).}
\bea
\label{rhotransf}
\rho ^i & = & R^{ii'} U^* \rho ^{i'} U^\dagger 
\no \\
\rho ^{ij} & = & R^{ii'} R^{jj'} U \rho ^{i'j'} U^\dagger
\no \\
\rho ^{ijk} & = & R^{ii'} R^{jj'} R^{kk'} U^* \rho ^{i'j'k'} U^\dagger
\eea
Under the above $SU(4)$ transformations, $\L_\sub$ is invariant,
even with varying gauge coupling $g(x^\pi)$.  In general,  however,
the interface Lagrangian $\L_I$ is not invariant because the interface couplings
transform non-trivially. Thus, $SU(4)$ maps one interface theory onto
another by mapping their respective interface couplings, 
\bea
\label{yztransf}
y^{i_1 ... i_n} & \to &  (y')^{i_1 ... i_n}
= 
R^{i_1  i_1'}... R^{i_n i_n'}  \, y^{i_1^\prime ... i_n^\prime}  
\no\\
z^{i_1 ... i_n} & \to & (z')^{i_1 ... i_n} 
=
R^{i_1 i_1'}... R^{i_n i_n'}  \, z^{i_1^\prime ... i_n^\prime} 
\eea
but leaving the gauge coupling unchanged. Under these combined transformations,
the reduced equations (\ref{aleq}) are  also invariant. Combining (\ref{rhotransf})
with (\ref{yztransf}), we have the following convenient formulas,
\bea
\label{global trans}
Y_2 \, \to \,  Y_2' = U Y_2 U^\dagger & \hskip 1in & Y_3 \, \to \, Y_3' = U Y_3 U^t
\no \\
Z_2 \, \to \,  Z_2'=U Z_2 U^\dagger & \hskip 1in & Z_3 \, \to \, Z_3' = U Z_3 U^t
\eea

\subsection{Further simplifying (3') and (5')}

To exhibit covariance of (\ref{aleq}), (\ref{seven'}) and (\ref{new35})
under local $x^\pi$-dependent $SU(4)$ transformations, 
and then  solve these equations, we shall begin by further simplifying the form
of (3') and (5'). To do so, we express $Y_3$ in terms of its coefficients $y_3 ^{ijk}$,
and $z_2^{ij}$ in terms of the matrix $Z_2$, both using (\ref{Y2Y3}),
\bea
\label{new35}
(3') & \hskip .1in &
      3 \left ( z_3^{ijk} - 8 y_3 ^{ijk} \right ) (\rho^k)^* \xi^*
     - \quart \left ( \rho^{ij} (Y_2+Z_2) - (Y_2+Z_2)  \rho^{ij} \right ) \xi 
      = 0
\no \\
(5') & \hskip .1in &
2 i y_1 \rho^i \xi 
+ \quart \left( (Y_2+Z_2) ^* \rho^i - \rho^i (Y_2+Z_2) \right) \xi  =0
\eea
using the relations,
\bea
Z_2 ^* \rho ^i - \rho ^i Z_2 & = & - 4 z_2 ^{ij} \rho ^j
\no \\
\rho ^{ij} Z_2 - Z_2 \rho ^{ij}
& = & 8 z_2 ^{ij} - 4 z_2 ^{jk} (\rho ^i)^* \rho ^k + 4 z_2 ^{ik} (\rho ^j)^* \rho^k
\no \\
24 y_3 ^{ijk} (\rho ^k)^* & = & 
\rho ^{ij} Y_3 + Y_3 (\rho ^{ij})^* 
- (\rho ^i)^* Y_3 ^* (\rho ^j)^* + (\rho ^j)^* Y_3 ^* (\rho ^i)^* 
\eea
The first two equations above were derived by contracting identities (\ref{rho2}) 
and (\ref{rho3})  with $z_2 ^{kl}$, and using (\ref{rho1}) to rearrange the second equation in the form above. The last equation was derived by contracting 
(\ref{rho4}) with $y_3 ^{klm}$, and making further simplifications using 
(\ref{rho1}) and (\ref{rho2}).

\subsection{Transformations under local $SU(4)$}

The fact that the interface couplings $y_2^{ij}$ and $z_2 ^{ij}$ may be 
regarded as arising from a $SU(4)$ background gauge field implies
that the Lagrangian $\L$, and the interface supersymmetry
conditions (\ref{aleq})  behave covariantly under local $x^\pi$-dependent 
$SU(4)$ transformations. 

\smallskip

Considering the equations of (\ref{aleq}), and their simplified forms in 
(\ref{seven'}) and (\ref{new35}), we see that only (4') involves a
derivative with respect to $x^\pi$. Thus, equations (1'), (2'), (3'),
(5') and (7') are actually covariant under {\sl local} $SU(4)$
transformations $\U$, and associated $\R$, whose action is  defined as follows,
\bea
\xi & \to & \xi ' = \U \xi
\no \\
\rho ^i & = & \R^{ii'} \U^* \rho ^{i'} \U^\dagger 
\eea
Covariance of (1') and (2') then requires the following transformation laws,
\bea
Y_3 & \to & \hskip .2in Y_3' = \U \, Y_3 \, \U^t
\no \\
z_1^{ij} & \to & (z'_1)^{ij} 
=
\R^{ii'} \, \R^{jj'}  \, z^{i'j'} _1
\eea
Covariance of the differential equation (4') requires that $Y_2$ transform
as a $SU(4)$ connection, while (5') requires that $Y_2+Z_2$ transform
homogeneously. As a result, 
\bea
Y_2 & \to & Y_2 ' = \U \, Y_2 \, \U^\dagger + \omega 
\hskip 1in \omega = {4 \over \p_\pi g} (\p _\pi \U) \, \U^\dagger
\no \\
Z_2 & \to & Z_2 ' = \U \, Z_2 \, \U^\dagger - \omega
\eea
and this induces the following transformation on $z_4 ^{ij}$,
\bea
z_4 ^{ij}  \to  (z_4 ')^{ij} = \R^{ii'} \R^{jj'} \left (
z_4 ^{ij} + \omega ^{ik} (z_1 ^{jk} + z_2 ^{jk})
+ \omega ^{jk} (z_1 ^{ik} + z_2 ^{ik}) - \omega ^{ik} \omega ^{jk} \right )
\eea
Under the above transformations, the equations (\ref{aleq}), and their simplified forms in (\ref{seven'}) and (\ref{new35}), are covariant.

\medskip

The $SU(4)$ gauge transformation, introduced above, is free of anomalies
and thus remains a viable map between full quantum theories. This may be
seen as follows. Its mixed anomalies with any semi-simple gauge current
vanish as both the $SU(4)$ and the gauge generators are traceless.
The triangle graph with three $SU(4)$ currents vanishes for kinematic
reasons. This may be seen from the fact that the only non-vanishing 
components of the ``gauge field" is $A_\pi (x^\pi) \sim y_2$ for which
$\epsilon ^{\mu \nu \rho \sigma } A_\nu \p_\rho A_\sigma $ and 
$\epsilon ^{\mu \nu \rho \sigma } A_\nu A_\rho A_\sigma $ vanish identically,
thereby canceling the perturbative anomaly.

\subsection{Choosing the gauge $Y_2=0$ and solving (4')}

The covariance under local $SU(4)$, established in the preceding section,
allows us to choose $\U(x^\pi)$ at will. Different choices of $\U$ will lead
to different values for the interface couplings, but to physically equivalent 
field theories. It will be convenient to choose
\bea
Y_2 =0
\eea
since the differential equation may then be easily solved,
\bea
\label{xisol}
\xi (x^\pi) & = & \xi _0 \, \beta \, e^{i \theta (x^\pi) /2}
\no \\
\theta (x^\pi ) & = & \theta (x_0 ^\pi) - 2 \int ^{x^\pi } _{x_0 ^\pi}
dx^\pi (\p_\pi g) \, y_1 (x^\pi)
\eea
Here, $\xi_0$ is a real, constant, Grassmann number; $\beta $ is 
a real constant 4-vector, left undetermined by equation (4'); and $x^\pi _0$ 
is an arbitrary reference point. It will be important in the sequel that $\xi$ 
maintains a direction independent of $x^\pi$.

\subsection{Diagonalizing  $Y_3$}

From equation (1') in (\ref{aleq}), it is manifest that rank($I- Y_3^* Y_3$)
provides an upper bound on the number of allowed interface supersymmetries.
In particular, when $Y_3=0$, as is the case for the CFT dual  
of the Janus solution, no interface supersymmetries will exist.
Of course, even when rank($I- Y_3^* Y_3$)$>1$, the solutions to
(1') will also have to satisfy (2'), (3') and (5'), which will impose
further restrictions. In this subsection, we use global $SU(4)$ 
transformations to solve (1') completely, as a function of $Y_3$.

\smallskip

We proceed as follows. We work first at one arbitrary point $x_0 ^\pi$.
Using only global $SU(4)$, we simplify (\ref{aleq}) by choosing a convenient 
frame to express the interface couplings. This simplification is carried out by first
diagonalizing $Y_3$ with the help of the following little 

\smallskip

\noindent
{\bf Lemma 1} ~ A complex symmetric matrix $Y_3=Y_3^t$ admits the following diagonalization,
\bea
\label{lemma1}
Y_3 = U D_3 U^t e^{  i \theta _0}
\eea
where $D_3$ is real diagonal, with non-negative entries, $\theta _0$ is a real 
phase,  and $U$ is special unitary, satisfying $U ^\dagger U=I$, and $\det \, U=1$.  
(When $\det (D_3)=0$, the angle $\theta _0$ may be absorbed in the special 
unitary rotation $U$.)
Note that the number of independent parameters on both sides are equal: 
20 for $Y_3$; 15 for $U$; 4 for $D_3$, and 1 for $\theta_0$.

\smallskip

We review the proof here. Using $Y_3^t=Y_3$, the matrix
$Y_3 Y_3^*= Y_3 Y_3 ^\dagger$ is seen to be Hermitian and non-negative, 
and may be diagonalized by a special unitary matrix $U$,
\bea
\label{diagY3}
Y_3 Y_3^* = U D_3 ^2 U^\dagger
\eea
Here, $D_3$ is real diagonal; its entries may be chosen to be non-negative. 
We prove the Lemma first when $D_3$ is invertible and has all distinct eigenvalues, 
in which case we have,  
$(D_3^{-1} U^\dagger Y_3) \, (D_3^{-1} U^\dagger Y_3)^\dagger = I$.
The matrix $D_3^{-1} U ^\dagger Y_3$ must be unitary (but not 
necessarily special unitary), and will be denoted by $U_1$,  so that 
$Y_3 = U D_3 U_1$. More generally, 
\bea
\label{diagY3a}
Y_3 = U \Lambda D_3 \Lambda^* U_1
\eea
where $\Lambda$ is an arbitrary diagonal special unitary matrix.
Equating the transpose of  (\ref{diagY3}), $Y_3 ^*Y_3 =  U^* D_3^2 U^t$, 
with  $Y_3 ^\dagger Y_3 = U_1 ^\dagger \Lambda D_3^2 \Lambda ^* U_1$,
obtained from (\ref{diagY3a}),  we find that the matrix $\Lambda ^* U_1 U^*$ 
must commute with $D_3^2$. Since the eigenvalues  of $D_3^2$ were assumed 
to be all distinct, $\Lambda ^* U_1 U^*$ must  be diagonal and unitary (though 
not necessarily special unitary). 
Since $\Lambda$ is an arbitrary special unitary matrix, upon which $U$ and 
$U_1$ do not depend, $\Lambda$ may be chosen so that 
$U_1 U^* = I e^{i \theta_0}$, which proves  Lemma 1 when $D_3$ 
is invertible and has all distinct eigenvalues.

\smallskip

When $D_3$ has zero and/or coincident eigenvalues, we make a small
perturbation on $Y_3$ continuously parametrized by $\ep$ which 
preserves $Y_3^t = Y_3$ but lifts any degeneracies and moves all 
eigenvalues of $D_3$ away from 0. The Lemma now holds for  
small $\ep$, of which $D_3$, $U$ and $\theta_0$ are continuous functions. 
The limit $\ep \to 0$ clearly exists since $U$ and $e^{i\theta_0}$ take 
values  in compact spaces. Finally, when $\det (D_3)=0$, the matrix $D_3$
has at least one zero on the diagonal, the presence of which may be used to 
rotate  away $\theta_0$ (a mechanism well-known to those familiar with the 
strong CP-problem).

\subsection{Solving equation (1')}

We start from the parametrization $Y_3 = U D_3 U^t e^{i \theta_0}$ proven
in Lemma 1. Equation (1') has non-vanishing solutions $\xi$ 
provided at least one eigenvalue of $D_3$ equals 1. Henceforth, we shall make 
this assumption, as we are only interested in interface theories with non-trivial supersymmetry.  Without loss of generality, we may choose the ordering of
the eigenvalues as follows,
\bea
\label{D3}
D_3 = {\rm diag} [1, a, b, c] \hskip 1in a,b,c \in {\bf R^+}
\eea
We introduce a basis of eigenvectors $\beta _a$, $a=1,2,3,4$ with 
components $(\beta _a)_b = \delta _{ab}$.
The existence of a non-vanishing solution $\xi$, whose $x^\pi$ dependence
was already determined in (\ref{xisol}) from solving (4'), fixes the 
$x^\pi$-dependence of $Y_3$.,
\bea
Y_3 (x^\pi) & = & U D_3 U^t \, e^{i \theta (x^\pi)}
\no \\
\xi (x^\pi ) & = & \xi _0  \beta \, e^{i \theta (x^\pi)/2} \hskip 1in \beta = U \beta _1
\eea
Since $\beta$ was $x^\pi$-independent by (\ref{xisol}), the $SU(4)$ matrix
$U$ must be $x^\pi$-independent as well. Therefore, $U$ may be rotated away
using only a global $SU(4)$ rotation. Whenever $a$, $b$, or $c$ equals 1,
more than a single solution will exist, with the same $x^\pi$-dependence
but with $\beta _1 $ replaced by a $\beta _a$, $a=2,3,4$.

\subsection{Solving equation (2')}

Equation (2') now simplifies to give
\bea
\label{red2}
e^{2i\theta } D_3 \rho^i \beta _1 = - (\rho^i)^* \beta _1 + z_1^{ij} (\rho^j)^* \beta _1
\eea
The following simple relations, deduced from the explicit form of 
the matrices $\rho ^i$ in (\ref{basisrho}) of Appendix A, 
\bea
\label{rhobeta}
&&
\rho ^1 \beta _1 =  i \, \beta _2 
\hskip 1in 
\rho ^3 \beta _1 = - i \beta _3
\hskip 1in 
\rho ^5 \beta _1 = + i \beta _4
\no \\ &&
\rho ^2 \beta _1 = + \beta _2 
\hskip 1in 
\rho ^4 \beta _1 = -  \beta _3
\hskip 1in 
\rho ^6 \beta _1 = + \beta _4
\eea
allow us to solve (\ref{red2}) by inspection, and we find,
\bea
\label{z3}
z_1^{11} = 1 - a \cos(2 \theta)
    \qquad
& z_1^{12} = z_1 ^{21}=  - a \sin(2 \theta) &
    \qquad
z_1^{22} =  1 + a \cos(2 \theta)
\no\\
z_1^{33} = 1 - b \cos(2 \theta)
    \qquad
& z_1^{34} = z^{43} _1 = - b \sin(2 \theta) &
    \qquad
z_1^{44} = 1 + b \cos(2 \theta)
\no\\
z_1^{55} =  1 - c \cos(2 \theta)
\qquad
& z_1^{56} = z_1 ^{65} = - c \sin(2 \theta) &
\qquad
z_1^{66} =  1 + c \cos(2 \theta)
\eea
with all other $z_1^{ij} = 0$.

\subsection{Solving equation (5')}

To solve (5') when $Y_2=0$, it will be convenient to use the 
original form of (5') in (\ref{aleq}), and then substitute the solution (\ref{xisol})
for $\xi$, 
\bea
2 i y_1 \rho^i \beta _1  - z_2^{ij} \rho^j \beta _1 =0
\eea
Using (\ref{rhobeta}), this equation may also be solved by inspection.
Clearly, all components of $z_2 ^{ij}$ must vanish, except 
\bea 
 z_2 ^{12} = z_2 ^{34} = z_2 ^{56}
= - z_2 ^{21} = - z_2 ^{43} = - z_2 ^{65} = -2 y_1 
\eea
It will be useful to recast this solution in the form of the matrix $Z_2$,
and we find,
\bea
\label{Z2sol}
Z_2 = - 4i y_1 \left ( I - 4 \beta _1 \otimes \beta _1 ^\dagger \right )
\eea

\subsection{Solving equation (3')}

With the help of the preceding results for $Y_2$, $Y_3$, and $Z_2$, equation
(3') becomes an equation for $z_3^{ijk}$. Indeed, substituting $Y_2=0$
and the solution of (\ref{Z2sol}) for $Z_2$ into (3') in (\ref{new35}), yields
the following expression,
\bea
\left ( z_3 ^{ijk} - 8 y_3 ^{ijk} \right ) (\rho ^k)^* \beta _1
= { 4 \over 3} i \, y_1 \, e^{i \theta } \left ( \rho ^{ij} \beta _1 
- \beta _1 \left ( \beta _1 ^\dagger \rho ^{ij} \beta _1 \right ) \right )
\eea
Both sides are manifestly orthogonal to $\beta _1$. Using again (\ref{rhobeta}), 
the projections onto $\beta _2, \beta _3, \beta _4$ may be computed as follows,
\bea
z_3 ^{ij1} - 8 y_3 ^{ij1} +i \left ( z_3 ^{ij2} - 8 y_3 ^{ij2} \right )
& = & - {4 \over 3} y_1 e^{i \theta } \beta _2 ^\dagger \rho ^{ij} \beta _1
\no \\
z_3 ^{ij3} - 8 y_3 ^{ij3} +i \left ( z_3 ^{ij4} - 8 y_3 ^{ij4} \right )
& = & + {4 \over 3} y_1 e^{i \theta } \beta _3 ^\dagger \rho ^{ij} \beta _1
\no \\
z_3 ^{ij5} - 8 y_3 ^{ij5} +i \left ( z_3 ^{ij6} - 8 y_3 ^{ij6} \right )
& = & - {4 \over 3} y_1 e^{i \theta } \beta _4 ^\dagger \rho ^{ij} \beta _1
\eea
Using the list of matrix elements in (\ref{rhobeta2}), these equations may
be solved by inspection, and we find $z_3 ^{12k} = z_3 ^{34k}=z_3 ^{56k}=0$
for all $k$. The remaining entries may be recast most easily in terms of the 
matrix $Z_3$ defined in (\ref{Y2Y3}),
\bea
Z_3 = 8 Y_3 + 32 i y_1 \, e^{  i \theta } \beta_1 \otimes \beta_1 ^t
\eea
This concludes the solution of the equations (\ref{aleq}).

\subsection{Summary of the solution}

For given gauge coupling $g(x^\pi)$, the solution to the supersymmetry constraint equations (\ref{summary}) 
is parameterized by the undetermined coefficient, $y_1 (x^\pi)$, the initial value of the
supersymmetry parameter phase, $\phi(x_0^\pi)$, two real, constant, Grassmann numbers $\xi_{0 \pm}$,
and the three unfixed eigenvalues of the real diagonal matrix $D_3 (x^\pi) = \mbox{diag}[1,a,b,c]$.
In terms of these parameters, the supersymmetry parameter is
\bea
\zeta (x^\pi) & = &  \lambda s_+ \otimes \xi_+ + i  \lambda ^* s_- \otimes \xi_-
\no\\
\xi_\pm (x^\pi) & = & \xi_{0 \pm} \, \beta _1 \, e^{i \theta (x^\pi) /2}
\no \\
\theta (x^\pi ) & = & \theta (x_{0}^\pi) - 2 \int ^{x^\pi } _{x_0 ^\pi}
dx^\pi (\p_\pi g) \, y_1 (x^\pi)
\no \\
(\beta _1) _a & = & \delta _{1 a}
\eea
where $s_+$ and $s_-$ are the two eigenvectors of $\gamma^\pi {\cal B}$ restricted to the + chirality subspace.
For every $\zeta$ that satisfies the supersymmetry equations we have two real Poincar\'e supersymmetries given by $\zeta$ and $\zeta^*$.  The cases with more than two supersymmetries will be discussed in the next section.
The fermionic coefficients are given by
\bea
\label{solution summary fermionic}
Y_2 & = & 0
\no\\
Y_3 & = & D_3 \, e^{i \theta}
\eea
The bosonic coefficients are given by $(\ref{z3})$ and
\bea
\label{solution summary bosonic}
Z_2 & = & - 4i y_1 \left ( I - 4 \beta _1 \otimes \beta _1 ^\dagger \right )
\no\\
Z_3 & = & 8 Y_3 + 32 i y_1 \, e^{  i \theta } \beta_1 \otimes \beta_1 ^t
\no\\
z_4^{ij} & = & - (z_1^{ik} + z_2^{ik})(z_1^{jk} + z_2^{jk})
\eea
Finally the modification to the supersymmetry transformation is given by
\bea
\chi^i = -  \left ( z_1^{ij} + z_2^{ij} \right ) (\rho^j)^* \gamma^\pi \B \zeta^*
\eea
Recall that we used both local and global $SU(4)$ transformations to 
simplify the form of $Y_2$ and $Y_3$, and to present the solution in 
the simple form given above. The general solution to the supersymmetry 
constraint equations may finally be recovered by undoing those $SU(4)$
transformations on the above solution, namely
by making local and global $SU(4)$ rotations on the above formulas.

\newpage

\section{Classifying interface supersymmetric theories}
\setcounter{equation}{0}

We now have all the tools available to give a complete classification of
the allowed Poincar\'e supersymmetries as a function of the gauge and 
interface couplings. Using Lemma 1, this classification proceeds
according to $r = {\rm rank}(Y_3 ^\dagger Y_3)$, and corresponds to
the following forms of $Y_3$ (after diagonalization by $U\in SU(4)$),
following Lemma 1,
\bea
\label{categories}
r=4 & \hskip 1in & Y_3 = e^{i \theta} \mbox{diag} [1,1,1,1 ] \no\\
r=3 & \hskip 1in & Y_3 = e^{i \theta} \mbox{diag} [1,1,1,c ] 
    \hskip 1in 0\leq c \not= 1 \no\\
r=2 & \hskip 1in & Y_3 = e^{i \theta} \mbox{diag} [1,1,b,c ] 
    \hskip 1in 0 \leq b \not= 1 \no \\
r=1 & \hskip 1in & Y_3 =e^{i \theta} \mbox{diag} [1,a,b,c]
    \hskip 1in 0 \leq a \not= 1
\eea
Within a category with fixed amount of supersymmetry (i.e. fixed $r$),
we shall principally be interested in the theory which has {\sl maximal
internal symmetry}, since other theories with the same amount of
supersymmetry but less internal symmetries may be viewed as
perturbations of the former by BPS operators that further break
the internal symmetry. 

\subsection{Solutions with extended interface supersymmetry }

Extended interface supersymmetry ($r >1$) will occur when solutions exist to the 
equations (\ref{aleq}), for given interface couplings, for at least two linearly independent unit vectors $\beta$. By $SU(4)$ symmetry, these vectors may be 
chosen to be $\beta _1$ and $\beta _2$, so that $Y_3$ is of the 
form  (\ref{categories}) with $r=2$. The cases with $r=3,4$ are contained
in this case by setting $b$ and possibly $c$ equal to 1. 

\smallskip

For a given interface theory, the  interface couplings $Z_2$ are fixed.
The simultaneous  solution for  $\beta _1$ and  $\beta _2$ will require
that the expression (\ref{Z2sol}) for $Z_2$ hold in terms of these two 
independent vectors, 
\bea
Z_2 & =  &
- 4i y_1 \left ( I - 4 \beta _1 \otimes \beta _1 ^\dagger \right )
\no \\
& = &
- 4i y_1 \left ( I - 4 \beta _2 \otimes \beta _2 ^\dagger \right )
\eea
These equations can hold only provided $y_1=0$, so that,
\bea
y_1 =  Y_2 = Z_2 =0 \hskip 1in Z_3 = 8 Y_3
\eea
which provides a considerable simplification of the solution, and renders
the supersymmetry parameter $\xi$ independent of $x^\pi$.
One can check that all of the supersymmetry equations in (\ref{aleq}) hold, 
except that (2') imposes the following consistency condition,
\bea
b = c \, e^{i 4 \theta} 
\eea
By construction of $D_3$, we have $b,c \geq 0$, so the solutions
fall into three distinct classes,
\bea
{\rm (I)} & \hskip .5in & b=c=1  \hskip .7in e^{4 i \theta }=1
    \hskip .81in SU(2) \times SU(2)
\no \\
{\rm (II)} & \hskip .5in & b=c =0 \hskip .7in  \theta ~ {\rm arbitrary}
     \hskip .59in SO(2) \times SU(2)
\no \\
{\rm (III)} & \hskip .5in & b=c \not= 0,1  \hskip .53in e^{4 i \theta }=1
    \hskip .83in SO(2) \times SO(2)
\eea
On the right side, we have listed the residual symmetry of the solution.
From these considerations, it is immediately clear that  the case $r=3$
does not support any solutions, since it would have $c \not= b =1$.

\subsection{The theory with $\N=4$ interface supersymmetry}

Case (I) corresponds to a theory with 8 conserved Poincar\'e supercharges, 
{\sl or $\N=4$ interface supersymmetry}, and $SU(2) \times SU(2) \sim SO(4)$ 
R-symmetry. Setting first the $SU(4)$ rotation matrix $U$ equal to 1, the interface Lagrangian is given by\footnote{The cases with  $\theta = \pm \pi/2, \pi$ are equivalent to $\theta = 0$, which is the choice  made here.}
\bea
\L_I &=& 
{(\partial_\pi g) \over g^3} \tr \bigg ( {i \over 2} \psi^t  \C \psi 
+ {i \over 2} \psi^\dagger \C \psi^*
- 4 i  \phi^2 [\phi^4,\phi^6] 
+ \p_\pi (\phi^2 \phi^2 + \phi^4 \phi^4 + \phi^6 \phi^6) \bigg)
\no\\ &&
- {(\partial_\pi g)^2 \over g^4} 2 \tr (\phi^2 \phi^2 + \phi^4 \phi^4 + \phi^6 \phi^6)
\eea
Using $(\ref{rescale})$, we can scale out the $(\p_\pi g)^2$ term by defining 
$\tilde \phi^i = \phi ^i / g^2$ for $i$ even.  Our conventions are as follows,
$\tilde \phi^i$ has $i \in \{2,4,6\}$ and $\phi^i$ has $i \in \{1,3,5\}$.
In terms of these fields the bulk Lagrangian is
\bea
\L_\sub &=& 
 - {1 \over 4 g^2} \tr \left ( F^{\mu \nu } F_{\mu \nu } \right ) 
- {1 \over 2 g^2} \tr (D^\mu \phi^i D_\mu \phi^i )
- {g^2 \over 2} \tr (D^\mu \tilde \phi^i D_\mu \tilde \phi^i )
\no \\ &&
+ { 1 \over 4 g^2} \tr ( [\phi^i, \phi^j ]  [\phi ^i, \phi ^j ] )
+ { g^2 \over 2 } \tr ( [\tilde \phi^i, \phi^j ]  [\tilde \phi ^i, \phi ^j ] )
+ { g^6 \over 4} \tr ( [\tilde \phi^i, \tilde \phi^j ]  [\tilde \phi ^i, \tilde \phi ^j ] )
\no \\ &&
- {i \over 2 g^2} \tr \left ( \bar \psi \gamma ^\mu D_\mu \psi \right )
+ {i \over 2 g^2} \tr \left ( D_\mu \bar \psi \gamma ^\mu \psi \right )
+ {1 \over 2 g^2} \tr \left ( \psi ^t \C  \rho^i [\phi ^i, \psi] 
+ \psi^\dagger \C (\rho^i)^*  [\phi ^i, \psi ^* ] \right )
\no \\ &&
+ {1 \over 2} \tr \left ( \psi ^t \C  \rho^i [\tilde \phi ^i, \psi] 
+ \psi^\dagger \C (\rho^i)^*  [\tilde \phi ^i, \psi ^* ] \right )
\eea
and the interface Lagrangian is
\bea
\L_I = 
{(\partial_\pi g) \over g^3} \tr \bigg (  {i \over 2} \psi^t  \C \psi 
+ {i \over 2} \psi^\dagger \C \psi^*
- 4 i  g^6 \tilde \phi^2 [\tilde \phi^4,\tilde \phi^6]  \bigg)
\eea
The $SU(2)\times SU(2) \subset SU(4)$ symmetry acts as follows. 
The first $SU(2)$ transforms the scalar triplet $(\phi ^1, \phi ^3, \phi ^5)$
in its 3-dimensional representation, while the second $SU(2)$
transforms the scalar triplet $ (\phi ^2, \phi ^4, \phi ^6)$. On the fermion
$\psi$, the group $SU(2) \times SU(2) \sim SO(4)$ acts in the (real)
4-dimensional representation of $SO(4)$. This theory clearly
admits a conformal limit, as all interface couplings proportional 
to $(\p_\pi g)^2$ have been eliminated. Therefore, the 8 Poincar\'e
supercharges are supplemented by 8 more conformal supercharges
to a total of 16 conformal supersymmetries.

\smallskip

Finally, we may restore the dependence on the $SU(4)$ rotation matrix
$U$ by including it in the form of $Y_3$ and $Z_3 = 8 Y_3$, which
gives us the general interface Lagrangian with $\N=4 $ interface supersymmetry,
\bea
\L_I = 
{(\partial_\pi g) \over g^3} \tr \bigg (  {i \over 2} \psi^t  \C U^* U^\dagger \psi 
+ {i \over 2} \psi^\dagger  \C U U^t \psi^*
+  i  g^6 \Re \, \tr \left ( \rho ^{ijk} U U^t \right )
\tilde \phi^i [\tilde \phi^j, \tilde \phi^k ]  \bigg)
\eea
where we have used (\ref{traces})  in re-expressing  the last term.
The space of interface theories is thus parametrized by the gauge coupling
and the interface couplings 
\bea
U \, \in \, {SU(4) \over SO(4)}
\eea
where $SO(4)$ acts on $U$ by right multiplication. Theories for different $U$ 
are physically equivalent, although described by a different set of couplings.

\subsection{The theories with $\N=2$ interface supersymmetry}

Cases (II) and (III) correspond to theories which have 4 conserved 
Poincar\'e supercharges, or {\sl  $\N=2$ interface supersymmetry}.
Case (II) has $SO(2) \times SU(2)$ R-symmetry, while (III) has a smaller 
internal symmetry $SO(2) \times SO(2)$. The theory corresponding to case 
(III) may be obtained from the theory corresponding to case (II) by adding 
interface operators that preserve the supersymmetry but further break the 
internal symmetry. We shall describe in detail only the theory with maximal 
internal symmetry, i.e. case (II). Since $D_3$ has zero eigenvalues, 
by Lemma 1, we may set $\theta =0$ without loss of generality, since the action
of this angle is equivalent to an $SU(4)$ rotation. Setting first the $SU(4)$
rotation matrix $U$ to 1, the interface Lagrangian for theory (II) is  given by,
\bea
\L_I & = & {(\p_\pi g) \over g^3} \tr \bigg ( 
{i \over 2} \psi^t \C D_3^{(2)} \psi
+ {i \over 2} \psi  ^ \dagger  \C D_3 ^{(2)} \psi^* 
+ 2 i \phi^2 [\phi^3, \phi^5]
- 2 i \phi^2 [\phi^4, \phi^6]
\no\\ && \hskip .6in
+ {1 \over 2} \p_\pi (\phi^i \phi^i + \phi^2 \phi^2 - \phi^1 \phi^1)
\bigg )
\no\\ &&
- {(\p_\pi g)^2 \over 2 g^4} \tr \left (\phi^i \phi^i + 3 \, \phi^2 \phi^2 - \phi^1 \phi^1
\right)
\eea
where $D_3 ^{(2)} = {\rm diag} [1,1,0,0]$. Using $(\ref{rescale})$, we can scale out the $(\p_{\pi} g)^2$ term by defining $\tilde \phi^2 = {\phi^2 \over g^2}$ and $\tilde \phi^i = {\phi^i \over g}$ for $i \in \{ 3, ... ,6 \}$.  In terms of $\tilde \phi^i$, the interface Lagrangian is
\bea
\L_I = {(\p_\pi g) \over g^3} \tr \left ( 
{i \over 2} \psi^t \C D_3^{(2)} \psi
+ {i \over 2} \psi  ^ \dagger  \C D_3 ^{(2)} \psi^* 
+ g^4 2 i \tilde \phi^2 [\tilde \phi^3, \tilde \phi^5]
- g^4 2 i \tilde \phi^2 [\tilde \phi^4, \tilde \phi^6] \right )
\eea
In this form, it is clear that these
interface theories admit a conformal limit, so that the 4 Poincar\'e supersymmetries
get enhanced with an extra 4 conformal supersymmetries.

\smallskip

Restoring the dependence of the 
interface couplings on the matrix $U$ gives the 
general interface Lagrangian with $\N=2$ interface supersymmetry,
\bea
\L_I & = &  {(\p_\pi g) \over g^3} \tr \bigg ( 
{i \over 2} \psi^t \C U^*D_3 ^{(2)} U^\dagger  \psi
+ {i \over 2} \psi  ^ \dagger  \C U D_3 ^{(2)} U^t \psi^* 
\no \\ && \hskip .6in
+  i  g^4 \Re \, \tr \left ( \rho ^{ijk} U D_3 ^{(2)} U^t \right )
\tilde \phi^i [\tilde \phi^j, \tilde \phi^k ]  \bigg )
\eea
The space of $\N=2$  interface theories is thus parametrized by the interface couplings 
\bea
U \, \in \, {SU(4) \over SO(2) \times SU(2)}
\eea
where $SO(2) \times SU(2)$ acts on $U$ by right multiplication, and 
leaves $D_3 ^{(2)}$ invariant. Theories for different $U$  are again physically 
equivalent, although described by different interface couplings.

\subsection{The theory with $\N=1$ interface supersymmetry}

Finally, the case $r=1$ in (\ref{categories}) corresponds to a family
of theories with 2 conserved Poincar\'e supercharges, {\sl or $\N=1$ 
interface supersymmetry}. Its maximal internal symmetry is $SU(3)$ 
attained in the case where $a=b=c=0$, while no internal symmetry 
remains when $a,b,c$ are all distinct. 
We shall describe in detail only the theory with maximal internal symmetry 
$SU(3)$ and $a=b=c=0$. Since $D_3$ has zero eigenvalues, by Lemma 1,
we may set $\theta =0$ without loss of generality. Using the results of $(\ref{conformal restrictions})$, 
we take $y_1 = 0$, so that we may scale out the $(\p_\pi g)^2$ term like before.
Setting first the $SU(4)$
rotation matrix to 1, the interface Lagrangian for  this theory is  given by,
\bea
\L_I &=& 
{(\partial_\pi g) \over g^3} \tr \bigg (
 {i \over 2}  \psi^t  \C D_3 ^{(1)} \psi 
 +  {i \over 2} \psi^\dagger \C D_3 ^{(1)} \psi^* \bigg) 
\\ &&
+ {(\partial_\pi g) \over g^3} \tr \bigg (
i  \phi^1 [\phi^3,\phi^6] 
+ i  \phi^1 [\phi^4,\phi^5]
+ i  \phi^2 [\phi^3,\phi^5] 
\no\\ && \hskip .8in
- i  \phi^2 [\phi^4,\phi^6]
+ {1 \over 2} \p_\pi (\phi^i \phi^i) \bigg)
- {(\partial_\pi g)^2 \over 2g^4} \tr \left ( \phi^i \phi^i \right )
\no 
\eea
where $D_3 ^{(1)}= {\rm diag} [1,0,0,0]$. We can scale out the $(\p_\pi g)^2$ 
term by defining 
$\tilde \phi^i = \phi ^i /g$.  In terms of $\tilde \phi^i$ the interface Lagrangian is
\bea
\L_I &=& 
{(\partial_\pi g) \over g^3} \tr \left (
{i \over 2}  \psi^t   \C D_3 ^{(1)} \psi 
+  {i \over 2} \psi^\dagger \C D_3 ^{(1)} \psi^* \right ) 
\\ &&
+ (\partial_\pi g)  \tr \bigg (
i \tilde \phi^1 [\tilde \phi^3,\tilde \phi^6] 
+ i \tilde \phi^1 [\tilde \phi^4,\tilde \phi^5]
+ i \tilde \phi^2 [\tilde \phi^3,\tilde \phi^5] 
- i \tilde \phi^2 [\tilde \phi^4,\tilde \phi^6] 
\bigg)
\no
\eea
In this formulation, it is clear that the theory admits a conformal limit, 
so that the 2 Poincar\'e supersymmetries get enhanced by an extra 
2 conformal supersymmetries. Notice, however, that the rescaling powers
of the gauge coupling $g$ are now different from those needed for the 
theories with $\N=4$ and $\N=2$ interface supersymmetry.

\smallskip

In terms of the complex fields 
\bea
\label{complexphi}
\Phi _1 = \tilde \phi ^1 + i \tilde \phi ^2
& \hskip 1in &
\Phi _1^* = \tilde \phi ^1 - i \tilde \phi ^2
\no \\
\Phi _2 = \tilde \phi ^3 + i \tilde \phi ^4
& \hskip 1in &
\Phi _2^* = \tilde \phi ^3 - i \tilde \phi ^4
\no \\
\Phi _3 = \tilde \phi ^5 + i \tilde \phi ^6
& \hskip 1in &
\Phi _3^* = \tilde \phi ^5 - i \tilde \phi ^6
\eea
the Lagrangian may be recast in a simplified form,
\bea
\L_I = 
 {(\partial_\pi g) \over g^3} \tr \left ( 
 {i \over 2}  \psi^t   \C D_3 ^{(1)} \psi 
 +  {i \over 2} \psi^\dagger \C D_3 ^{(1)} \psi^*
+ {g^3 \over 2} \Phi _1 [\Phi _2, \Phi _3] 
- {g^3 \over 2}  \Phi _1 ^* [\Phi _2 ^* , \Phi _3 ^*] \right )
\eea
In this form, the $SU(3)$ symmetry is manifest on the scalar fields
as well as on the fermions, and rotates the fields $\Phi_1, \Phi _2$, 
and $\Phi_3$ into one another. 

\smallskip

Restoring the dependence of the interface couplings on the matrix $U$ gives the 
general interface Lagrangian with $\N=1$ interface supersymmetry,
\bea
\L_I & = &  {(\p_\pi g) \over g^3} \tr \bigg ( 
{i \over 2} \psi^t \C U^*D_3 ^{(1)} U^\dagger  \psi
+ {i \over 2} \psi  ^ \dagger  \C U D_3 ^{(1)} U^t \psi^* 
\no \\ && \hskip .6in
+  i  g^3 \Re \, \tr \left ( \rho ^{ijk} U D_3 ^{(1)} U^t \right )
\tilde \phi^i [\tilde \phi^j, \tilde \phi^k ]  \bigg )
\eea
The space of physically equivalent $\N=1$  interface theories is thus 
parameterized by the interface couplings 
\bea
U \, \in \, {SU(4) \over SU(3)}
\eea
where $SU(3)$ acts on $U$ by right multiplication, and leaves 
$D_3 ^{(1)}$ invariant. Of this 7-dimensional space of interface 
couplings, 6 may be absorbed by taking suitably modified
linear combinations for the definition of the fields $\Phi_1$,
$\Phi_2$, and $\Phi_3$ in (\ref{complexphi}). The remaining
diagonal generator of $SU(4)$ leaves the following angle,
\bea
\L_I & = & 
 {(\partial_\pi g) \over g^3} \tr \bigg ( 
 {i \over 2}  e^{ + 3 i \theta}\psi^t   \C D_3 ^{(1)} \psi 
 +  {i \over 2} e^{-3i \theta} \psi^\dagger \C D_3 ^{(1)} \psi^*
 \no \\ && \hskip .5in 
+ {g^3 \over 2} e^{+ 3i \theta} \Phi _1 [\Phi _2, \Phi _3] 
- {g^3 \over 2}  e^{-3 i \theta} \Phi _1 ^* [\Phi _2 ^* , \Phi _3 ^*] \bigg )
\eea

\smallskip

In \cite{Clark:2004sb} this Lagrangian was constructed using the language of $\N=1$ theories.  It was necessary to use different scalings for the chiral and vector multiplets.  We now see that the origin of this lies in eliminating the $z_4$ 
term proportional to $(\p_\pi g)^2$ in order to guarantee the existence of a conformal limit.  Alternatively, one may keep the same scaling for the chiral and vector multiplets but obscure conformal invariance.

\subsection{Summary of symmetry considerations}

In the interface theory, the $SU(4)$ transformations of the original $\N=4$ 
super Yang-Mills naturally fall into three distinct classes.  

\smallskip

Transformations in the first class form the subgroup $G$ of $SU(4)$ which 
leaves $D_3$ of (\ref{lemma1}) and the supersymmetry, or $\xi_\pm$, invariant.\footnote{In the case of more than one supersymmetry, we take 
$G$ to leave invariant all of the supersymmetries as well as $D_3$.}  
It follows that $G$ will leave all other interface couplings invariant, thus 
yielding an {\sl internal} symmetry of the full interface theory.  To see this 
note that invariance of $Z_2$, $Z_3$, $Y_2$ and $Y_3$ follows from $(\ref{solution summary fermionic})$, $(\ref{solution summary bosonic})$ and the transformation rules $(\ref{global trans})$; $z_1^{ij}$ 
is completely determined by $Y_3$ and $\xi_\pm$ which are invariant; 
invariance of $z_4^{ij}$ follows from invariance of $z_1^{ij}$ and $z_2^{ij}$.
These groups are $G=SU(3)$, $G=SU(2)$ and the identity group $G=\{I\}$ 
for the theories with 2, 4, and 8 Poincar\'e supercharges respectively.

\smallskip

Transformations in the second class form a subgroup of $SU(4)$ which leaves 
the interface couplings invariant, but transforms the supersymmetries into 
one another, thus yielding the R-symmetry of the full interface theory.  
The R-symmetry group turns out to be $SO(n)$, where $n$ is the number of linearly independent $\xi$ that satisfy the supersymmetry constraint equations 
$(\ref{aleq})$. The space-time symmetries may then be conveniently packaged 
in the supergroup $OSP(2,2|n)$.

\smallskip

Transformations in the third class are dualities generated by $U \in SU(4)$ 
which do not lie in $G$ or $SO(n)$.   These transformations change the interface couplings, and thus provide a map between interface theories related by $SU(4)$ rotations.  Since the supersymmetry constraint equations are covariant under $SU(4)$, these different interface theories will have the same amount of supersymmetry.  Such transformations are parameterized by the coset 
\bea
{ SU(4) \over  G \times SO(n)}
\eea 
and relate different, but physically equivalent theories.  This is analogous to Montonen-Olive duality, which states that two 
different theories related by $SL(2,{\bf Z})$ transformations are physically 
equivalent.  A special role is played by transformations 
$U \in SU(4) /( G \times SO(n))$ which do not change the embedding of 
$G \times SO(n)$.  These transformations form the subgroup $H$
of $SU(4)$ which commutes with  $G \times SO(n)$.  

\smallskip

For the $AdS$ dual, not all of the $SU(4)$ transformations can be implemented 
in a useful way.  First, one can choose an Ansatz that is invariant under 
$G \times SO(n)$.  Once this is done, the embedding of $G \times SO(n)$ is fixed and the only useful transformations of $SU(4)$ that can be implemented on the 
$G \times SO(n)$-invariant Ansatz are those in $H$.  Since we did not include 
$\tr ( F \tilde F )$ or $\L_{CS}$, given in $(\ref{interface couplings})$, 
in our analysis, our theory cannot exhibit Montonen-Olive duality.  
We expect that upon including these terms, we should recover Montonen-Olive duality, which degenerates to $SL(2,{\bf R})$ in the supergravity 
limit. This gives the following symmetry Ansatz for the $AdS$ dual
\bea
SO(2,3) \times SO(n) \times G \times H \times SL(2,{\bf R})
\eea
Based on these symmetry considerations for the Ansatz of the AdS dual,
an explicit ten dimensional supergravity solution is found in \cite{edjemg}, 
with $G = SU(3)$, and the solution generating symmetries 
$U(1)_\beta  \times SL(2,{\bf R})$.

\newpage

\section{Conformal symmetry}
\setcounter{equation}{0}

We now analyze under what conditions conformal symmetry may be recovered.  
First note that the gauge coupling needs to  be invariant under a scale transformation.  One way to achieve this is to take the gauge coupling to be a 
step function.  However, the $z_4$ term in the interface Lagrangian $(\ref{Linterface})$ is proportional to $(\p_\pi g)^2$ and does not have a 
well-defined limit.  In some cases it is possible to eliminate $z_4$ by rescaling 
the scalar fields.   An example of this was already shown in \cite{Clark:2004sb}.  
We examine the conditions necessary for such a rescaling to work in general.  
We define rescaled scalar fields $\tilde \phi^i$ by $\phi^i = g^{n_i} \tilde \phi^i$, 
where $n_i$ is the real scaling exponent which is to be solved for. 
The relevant part of the Lagrangian is 
\bea
\L_1 = 
- {1 \over 2 g^2} \tr \left ( D_\mu \phi^i D^\mu \phi^i \right ) 
+ {(\partial_\pi g) \over 2g^3} z_1^{ij} \partial_\pi \tr \left ( \phi^i \phi^j \right ) 
+ {(\partial_\pi g)^2 \over 2g^4} z_4^{ij} \tr \left ( \phi^i \phi^j \right ) 
\eea
In terms of rescaled fields the Lagrangian becomes
\bea
\L_1 &=& 
-  {g^{2n_i} \over 2 g^2} \, \tr \left ( D_\mu \tilde \phi ^i D^\mu \tilde \phi ^i \right )
- {g^{n_i+n_j}  \over 2 g^3} 
\left ( n_i \delta _{ij} - z_1 ^{ij} \right )   (\p _\pi g) 
\, \p_\pi  \tr \left ( \tilde \phi ^i \tilde \phi ^j \right )
\no \\ &&
- {g^{n_i+n_j}  \over 2 g^4}
\left ( n_i ^2 \delta _{ij} - (n_i + n_j) z_1 ^{ij} -z _4 ^{ij} \right )  (\p _\pi g)^2
\, \tr \left ( \tilde \phi ^i \tilde \phi ^j \right )
\eea
The conformal limit will exist, provided there exists a  rescaling  
$\phi ^i = g^{n_i} \tilde \phi ^i$ such that the term proportional to 
$(\p _\pi g)^2$ vanishes; this can be achieved when
\bea
n_i ^2 \delta _{ij} - (n_i + n_j) z_1 ^{ij} -z _4 ^{ij} =0
\eea
For $i = j$, we obtain  6 equations which determine the 
exponents $n_i$, as follows,
\bea
\label{rescale}
0 & = & - n_i^2 + 2 n_i z_1^{ii} + z_4^{ii} 
\no \\
n_i & = & z_1^{ii} \pm \sqrt{ (z_1^{ii})^2 + z_4^{ii}}
\eea
Using the explicit form for $z_4^{ij}$, obtained from the supersymmetry
condition (\ref{seven'}), we deduce a simple expression for the argument 
of the square root in the formula for $n_i$,
\bea
(z_1^{ii})^2 + z_4 ^{ii} 
=
- \sum _{k\not= i} \left ( z_1 ^{ik} + z_2 ^{ik} \right )^2
\eea
Reality of $n_i$ requires these arguments to be non-negative,
which can hold only when,
\bea
\label{conformal restrictions}
z_1 ^{ik} = z_2 ^{ik} =0 \hskip 1in i\not= k
\eea
These results imply that $y_1 = 0$ along with either $\sin (2 \theta)=0$
and/or $a=b=c=0$.

\newpage

\section{Comparing with duals to brane configurations}
\setcounter{equation}{0}

As was remarked in the Introduction, no compelling brane configuration candidate 
is available yet which naturally reduces to the non-supersymmetric Janus solution
in the near-horizon limit. It may be hoped that, with the additional restrictions
provided by supersymmetry, a comparison between candidate brane configurations
and interface dual Yang-Mills theories will be considerably facilitated.
The more supersymmetry, the more restrictive the comparison will be.

\smallskip

The interface Yang-Mills theory with maximal supersymmetry
has $\N=4$ interface supersymmetry, or 8 real Poincar\'e supercharges, and $SU(2)\times SU(2)$ internal symmetry. Remarkably, there exists a brane 
configuration which has precisely those symmetries, and yet does not produce 
our $\N=4$ interface theory in the near-horizon limit. Indeed, a configuration with intersecting D3 and D5 branes  has 8 real Poincar\'e supercharges, which is 
1/4 of the maximal 32 supersymmetries of the theory. 
This configuration also has $SU(2) \times SU(2) \sim SO(4)$ internal symmetry. 
In the near-horizon limit, the Poincar\'e symmetry gets enhanced to 
conformal $SO(2,3)$, and the 8 Poincar\'e supercharges are supplemented
with 8 special conformal supercharges, just as is the case in the conformal 
limit of the $\N=4$ interface supersymmetry theory. 
Thus, the symmetries of the D3-D5 brane system and of our $\N=4$ 
interface theory match precisely. 

\smallskip

And yet, the supersymmetry multiplet content is different, because the D3-D5 
system contains open $3-5$ strings stretching from the D3 to the D5 brane. 
Therefore, the CFT dual to the D3-D5 system is expected to include gauge degrees of freedom localized on the interface. In \cite{DeWolfe:2001pq}, 
a 2+1-dimensional hypermultiplet, localized on the brane, in the fundamental representation of the gauge group $SU(N)$ was proposed for such defect 
degrees of freedom. Ample evidence was presented there that this scenario
is viable, at least in the approximation where the D5 brane is treated 
as a ``probe", and the back reaction of the D5 on the system of D3
branes is ignored \cite{Karch:2000gx,Aharony:2003qf,Erdmenger:2002ex,Yamaguchi:2003ay}. In our $\N=4$ interface theory, however, such extra 
degrees of freedom are absent. If our $\N=4$ interface theory is to have any 
relation with the D3-D5 brane system, some mechanism must exist
that decouples the $3-5$ strings from the CFT dual theory.\footnote{An example for such a mechnism in a different system was
discoverd in \cite{Itzhaki:2005tu}.}  We hope to investigate these
questions in the future.

\newpage

\vskip .5in

\noindent{\Large \bf Acknowledgments}

\vskip .1in

\noindent 
It is a pleasure to acknowledge helpful conversations with  Iosif Bena, 
Per Kraus and Norisuke Sakai.
This work was supported in part by National Science Foundation (NSF)
grant PHY-04-56200.
ED is grateful to the Kavli Institute for Theoretical Physics (KITP)
for their hospitality and support under NSF grant PHY-99-07949.
MG is grateful to the Harvard Particle Theory group for hospitality
while this work was being completed.

\newpage

\begin{appendix}

\section{Algebra of $SU(4)$ Clebsch-Gordan coefficients}
\setcounter{equation}{0}

The canonical map between $SU(4)$ and $SO(6)$ is realized in terms
of the Clifford algebra of $SO(6)$, given by the $\gamma^i_{(6)}$-matrices,
which obey
\bea
\{ \gamma ^i _{(6)}, \gamma ^j _{(6)} \} = 2 \delta ^{ij} I_8 
\hskip 1in
i,j=1,\cdots, 6
\eea
The chirality matrix is defined by 
$\gamma _{(6)} \equiv i \gamma ^1 _{(6)} \gamma ^2 _{(6)} 
    \cdots  \gamma ^6 _{(6)}$ 
and satisfies $(\gamma _{(6)})^2 = I_8$. The charge conjugation matrix $\C_{(6)}$ 
is defined by $\C_{(6)} \gamma ^i _{(6)} \C_{(6)} ^{-1} = - (\gamma ^i _{(6)} )^t$, 
and satisfies  the relations $\C^t _{(6)} = \C_{(6)}^{-1} = \C_{(6)}^* = \C_{(6)}$. 
We choose the chiral basis of $\gamma$-matrices, where,  
\bea
\gamma _{(6)} = \left ( \matrix{I_4 & 0 \cr 0 & - I_4 \cr} \right )
\eea
and $\C_{(6)} \gamma _{(6)} = - \gamma _{(6)} \C_{(6)}$. 
The chiral restriction of the Dirac matrices then defines the matrices $\rho^i$
as follows, 
\bea
\label{rhogamma}
\C_{(6)} \gamma ^i _{(6)} & = & 
\left ( \matrix{\rho ^i  & 0 \cr 0 & - (\rho^i)^* \cr} \right )
\no \\
\gamma ^i _{(6)} \C_{(6)}  & = & 
\left ( \matrix{- (\rho ^i)^*  & 0 \cr 0 & \rho^i \cr} \right )
\eea
We shall also need the chiral restriction of Clifford algebra generators of higher rank $SO(6)$-tensors of the form 
$\gamma ^{i_1 \cdots i_p} _{(6)} \equiv \gamma _{(6)} ^{[i_1} 
\cdots \gamma _{(6)} ^{i_p]}$. 
They are given as follows, 
\bea
\gamma _{(6)} ^{ij} 
=
\left ( \matrix{\rho ^{ij}  & 0 \cr 0 &  (\rho^{ij})^* \cr} \right ) 
\hskip .25in
& \hskip .8in &
\rho ^{ij} \equiv  - (\rho ^{[i} )^* \rho ^{j]} 
\no \\
\C_{(6)} \gamma ^{ijk} _{(6)} 
= 
\left ( \matrix{\rho ^{ijk}  & 0 \cr 0 &  - (\rho^{ijk})^* \cr} \right )
& \hskip .8in &
\rho ^{ijk} \equiv   - \rho ^{[i}  (\rho ^j)^* \rho ^{k]}
\no \\
\gamma ^{ijk} _{(6)} \C_{(6)} 
= 
\left ( \matrix{- (\rho ^{ijk})^*  & 0 \cr 0 &  \rho^{ijk} \cr} \right )
& \hskip .8in &
\eea
Square brackets $[ \, ]$ imposes anti-symmetrization on whichever set of indices 
occurs inside. In view of the definition of charge conjugation, the matrices $\rho ^i$ are anti-symmetric, which implies the following properties,
\bea
\label{conjugation}
(\rho ^i )^t & = & - \rho ^i
\hskip 1.07in 
{\bf 4} \otimes {\bf 4} ~ \quad \to \quad {\bf 6}
\no \\
(\rho ^{ij})^\dagger & = & - \rho ^{ij}
\hskip 1in 
{\bf 4} \otimes {\bf \overline{4}} \quad ~ \to \quad {\bf 15}
\no \\
(\rho ^{ijk} )^t & = & + \rho ^{ijk}
\hskip .96in 
{\bf 4} \otimes {\bf 4} \, \quad \to \quad {\bf 10}
\eea
These matrices provide the Clebsch-Gordan coefficients for the $SU(4)$ 
tensor product decompositions involving the ${\bf 4}$ and $  {\bf \overline{4}} $
representations, listed above in the right column.

\subsection{The algebra of the $\rho$-matrices}

The standard Clifford algebra readily provides relations
between the products of $\rho$-matrices in various combinations. 
The following Clifford relations,
\bea
\left \{ \gamma ^i , \gamma ^j \right \} 
& = & 2 \delta ^{ij} I
\no \\
\left [ \gamma ^{ij} , \gamma ^k \right  ] 
& = & 2 \delta ^{jk} \gamma ^i 
- 2 \delta ^{ik} \gamma ^j 
\no \\
\left [ \gamma ^{ij} , \gamma ^{kl} \right  ] 
& = & 2 \delta ^{jk} \gamma ^{il} 
+ 2 \delta ^{jl} \gamma ^{ki} 
- ( i \leftrightarrow j)
\no \\
\left [ \gamma ^{ij} , \gamma ^{klm} \right  ] 
& = & 2 \delta ^{jk} \gamma ^{ilm} 
+ 2 \delta ^{jl} \gamma ^{kim} 
+ 2 \delta ^{jm} \gamma ^{kli} 
- ( i \leftrightarrow j)
\no \\
\left \{ \gamma ^i , \gamma ^{klm} \right  \}
& = & 2 \delta ^{ik} \gamma ^{lm} 
+ 2 \delta ^{il} \gamma ^{mk} 
+ 2 \delta ^{im} \gamma ^{kl} 
\eea
translate to the following equations for the $\rho$-matrices and their complex conjugates,
\bea
\label{rho1}
(\rho  ^i)^* \rho ^j + (\rho ^j)^* \rho^i 
& = & - 2 \delta ^{ij} I
 \\
\label{rho2}
(\rho  ^{ij})^* \rho ^k - \rho ^k \rho^{ij}
& = &  
2 \delta ^{jk} \rho ^i - 2 \delta ^{ik} \rho ^j
 \\
\label{rho3}
\left [ \rho ^{ij}, \rho ^{kl} \right ]
& = &
2 \delta ^{jk} \rho ^{il} + 2 \delta ^{jl} \rho ^{ki} 
- ( i \leftrightarrow j)
\\
\label{rho4}
(\rho  ^{ij})^* \rho ^{klm} - \rho ^{klm} \rho^{ij}
& = &  
2 \delta ^{jk} \rho ^{ilm} + 2 \delta ^{jl} \rho ^{kim} + 2 \delta ^{jm} \rho ^{kii}
- ( i \leftrightarrow j)
\\
\label{rho5}
(\rho  ^i)^* \rho ^{klm} + (\rho ^{klm})^* \rho^i
& = &  
-2 \delta ^{ik} \rho ^{lm} - 2 \delta ^{il} \rho ^{mk} - 2 \delta ^{im} \rho ^{kl}
\eea
These relations provide the Clebsch-Gordan coefficients for tensor 
products involving the {\bf 6}, {\bf 15}, {\bf 10}, and ${\bf \overline{10}}$ representations.

\smallskip

The following relations will also be useful,
\bea
\label{traces}
\tr \left ( \rho ^i (\rho ^{i'})^* \right ) & = & - 4 \delta ^{ii'}
\no \\
\tr \left ( \rho ^{ijk} (\rho ^{i'j'k'} )^* \right ) & = & + 4 \delta ^{i[i'}
\delta ^{j j'} \delta {k k']}
\eea
where the bracket $[ ~ ]$ indicates anti-symmetrization of the primed indices only.

\subsection{Explicit form in standard basis}

The standard basis of $\gamma _{(6)} ^i$
matrices yields the following convenient basis of $\rho^i$ matrices,
\bea
\label{basisrho}
\rho ^1 = \sigma _3 \otimes \sigma _2
& \hskip 1in &
\rho ^2 = -i I \otimes \sigma _2
\no \\
\rho ^3 =  - \sigma _2 \otimes I
& \hskip 1in &
\rho ^4 = i \sigma _2 \otimes \sigma _3
\no \\
\rho ^5 = \sigma _1 \otimes \sigma _2
& \hskip 1in &
\rho ^6 = -i \sigma _2 \otimes \sigma _1
\eea
In this basis, we have simple complex conjugation relations, (in addition
to the general relations of (\ref{conjugation})),
\bea
(\rho ^i)^* & = & (-1)^i \rho ^i
\\
(\rho ^{ij})^* & = & (-1) ^{i+j} \rho ^{ij}
\hskip 1in 
\rho ^{ij} = - (-1)^i \rho ^i \rho ^j
\hskip .68in j\not=i
\no \\
(\rho ^{ijk})^* & = & (-1)^{i+j+k} \rho ^{ijk}
\hskip .7in 
\rho ^{ijk} = - (-1)^j \rho ^i \rho ^j \rho ^k
\hskip .5in k \not=i,j
\no \eea
while the square of each $\rho^i$ matrix is given by $(\rho ^i)^2 = - (-1)^i$.
These formulas facilitate the evaluation, in this basis, 
the matrices $\rho ^{ij}$, 
\bea
\label{basisrhoij}
\rho ^{12} =  - i \sigma _3 \otimes I 
    & \qquad \rho ^{23} =   - i \sigma _2 \otimes \sigma _2 
    & \qquad \rho ^{35} =   i \sigma _3 \otimes \sigma _2
\no \\
\rho ^{13} =   i \sigma _1 \otimes \sigma _2 
    & \qquad \rho ^{24} =  - i \sigma _2 \otimes \sigma _1
    & \qquad \rho ^{36} =  i I \otimes \sigma _1 
\no \\
\rho ^{14} =  i \sigma _1 \otimes \sigma _1 
    & \qquad \rho ^{25} = ~ \,  i \sigma _1 \otimes I 
    & \qquad \rho ^{45} =   i \sigma _3 \otimes \sigma _1 
\no \\
\rho ^{15} =   i \sigma _2 \otimes I ~
    & \qquad \rho ^{26} = - i \sigma _2 \otimes \sigma _3 
    & \qquad \rho ^{46} =  - i I \otimes \sigma _2 
\no \\
\rho ^{16} =   i \sigma _1 \otimes \sigma _3 
    & \qquad \rho ^{34} =   - i I \otimes \sigma _3
    & \qquad \rho ^{56} = - i \sigma _3 \otimes \sigma _3 
\eea
and the matrices $\rho ^{ijk}$, 
\bea
\rho ^{123} = + i \rho ^{456} = \, - \sigma _1 \otimes I \,
    & \hskip .5in &
    \rho ^{135} =-i \rho ^{246} =  i I \, \otimes I \, 
\no \\
\rho ^{124} = -i \rho ^{356} = ~i \sigma _1 \otimes \sigma _3
    & \hskip .5in &
    \rho ^{136} = +i \rho ^{245} = \sigma _3 \otimes \sigma _3
\no \\
\rho ^{125} = +i \rho ^{346} = - \sigma _2 \otimes \sigma _2
    & \hskip .5in &
    \rho ^{145} = +i \rho ^{236} = \, I \, \otimes \sigma _3
\no \\
\rho ^{126} = -i \rho ^{345} = - i \sigma _1 \otimes \sigma _1
    & \hskip .5in &
    \rho ^{146} = -i \rho ^{235} =  -i \sigma _3 \otimes I
\no \\
\rho ^{134} = +i \rho ^{256} = ~~~ \sigma _3 \otimes \sigma _1
    & \hskip .5in &
    \rho ^{156} = -i \rho ^{234} = I \otimes \sigma _1
\eea
By inspection, it is now manifest that $\rho ^i$, $\rho ^{ij}$ and $\rho ^{ijk}$
provide a basis respectively for complex antisymmetric, antihermitian, and 
complex symmetric traceless $4 \times 4$ matrices, confirming the 
Clebsch-Gordan maps of (\ref{conjugation}).

\smallskip

Finally, the following matrix elements will be useful when solving (3'),
\bea
\label{rhobeta2}
\rho ^{12} \beta _1 = -i \beta _1 
\qquad  & \rho ^{13} \beta _1 = - \beta _4
\qquad  & \rho ^{14} \beta _1 = i \beta _4
\no \\
\rho ^{15} \beta _1 = - \beta _3 
\qquad  & \rho ^{16} \beta _1 = i \beta _3
\qquad  & \rho ^{23} \beta _1 = i \beta _4
\no \\
\rho ^{24} \beta _1 =  + \beta _4 
\qquad  & \rho ^{25} \beta _1 = i \beta _3
\qquad  & \rho ^{26} \beta _1 = + \beta _3
\no \\
\rho ^{34} \beta _1 = -i \beta _1 
\qquad  & \rho ^{35} \beta _1 = - \beta _2
\qquad  & \rho ^{36} \beta _1 = i \beta _2
\no \\
\rho ^{45} \beta _1 = i \beta _2 
\qquad  & \rho ^{46} \beta _1 = + \beta _2
\qquad  & \rho ^{56} \beta _1 = -i \beta _1
\eea

\end{appendix}

\newpage

\end{document}